\documentclass[a4paper,11pt]{article}
\pdfoutput=1
\usepackage[utf8x]{inputenc}
\usepackage{jheppub,color,subfigure,slashed,hyperref}

\hypersetup{bookmarks=true,unicode=true,pdftoolbar=true,pdfmenubar=true,
	pdffitwindow=false,pdfstartview={FitH},pdftitle={VBF},
	pdfauthor={J.~Araz,~S.~Banerjee,~R.S.~Gupta,~M.~Spannowsky}, pdfsubject={},
	pdfcreator={},pdfproducer={}, pdfkeywords={},
	pdfnewwindow=true,colorlinks=true,
	linkcolor=blue,citecolor=magenta,filecolor=magenta,urlcolor=cyan}

\newcommand{\bea}{\begin{eqnarray}}
\newcommand{\eea}{\end{eqnarray}}
\newcommand{\eq}[1]{eq.~(\ref{#1})}

\newcommand{\nn}{\nonumber}

\def\stw{s_{\theta_W}}
\def\ctw{c_{\theta_W}}

\def\lra#1{\overset{\text{\scriptsize$\leftrightarrow$}}{#1}}

\makeatletter
\def\section{\@startsection {section}{1}{\z@}{-3.5ex plus -1ex minus
 -.2ex}{2.3ex plus .2ex}{\large\bf}}
\def\subsection{\@startsection{subsection}{2}{\z@}{-3.25ex plus -1ex
minus -.2ex}{1.5ex plus .2ex}{\normalsize\bf}}
\makeatother
\makeatletter

\@addtoreset{equation}{section}

\makeatother

\newcommand{\madanalysis}{{\sc MadAnalysis 5}}
\newcommand{\madgraph}{{\sc MadGraph}}
\newcommand{\pythia}{{\sc Pythia}}

\newcommand{\fastjet}{{\sc FastJet}}

\preprint{IPPP/20/52, CERN-TH-2020-186}

\begin{document}

\title{Precision SMEFT bounds from the VBF Higgs at high transverse momentum}
\author[a]{Jack Y. Araz,}
\author[b]{Shankha Banerjee,}
\author[a]{Rick~S.~Gupta,}
\author[a]{Michael~Spannowsky}
\affiliation[a]{Institute for Particle Physics Phenomenology,\\Durham University, South Road, Durham, DH1 3LE,}
\affiliation[b]{CERN, Theoretical Physics Department, CH-1211 Geneva 23, Switzerland }

\emailAdd{jack.araz@durham.ac.uk}
\emailAdd{shankha.banerjee@cern.ch}
\emailAdd{sandeepan.gupta@durham.ac.uk}
\emailAdd{michael.spannowsky@durham.ac.uk}
\date{\today}

\abstract{We study the production of Higgs bosons at high transverse momenta via vector-boson fusion (VBF) in the Standard Model Effective Field Theory (SMEFT). We find that contributions from four independent operator combinations dominate in this limit. These are the same `high energy primaries' that control high energy diboson processes, including Higgs-strahlung.  We perform detailed collider simulations for the diphoton decay mode of the Higgs boson as well as the three final states arising from the ditau channel. Using the quadratic growth of the SMEFT contributions relative to the Standard Model (SM) contribution, we project very stringent bounds on these operators that far surpass the corresponding bounds from the LEP  experiment.}

\maketitle

	\date{\today}
	
\section{Introduction}\label{sec:intro}
	
In the absence of any evidence for new physics at the  Large Hadron Collider (LHC), the Standard Model Effective Field Theory (SMEFT) is an efficient parametrisation for heavy new physics beyond the reach of the LHC. The effective field theory (EFT) formalism has, in fact,  become the standard framework to precision physics at the LHC~\cite{Buchmuller:1985jz, Giudice:2007fh, Grzadkowski:2010es, Gupta:2011be,Gupta:2012mi,Banerjee:2012xc, Gupta:2012fy, Banerjee:2013apa, Gupta:2013zza,Elias-Miro:2013eta, Contino:2013kra, Falkowski:2014tna,Englert:2014cva, Gupta:2014rxa, Amar:2014fpa, Buschmann:2014sia, Craig:2014una, Ellis:2014dva, Ellis:2014jta, Banerjee:2015bla, Englert:2015hrx, Ghosh:2015gpa, Degrande:2016dqg,Cohen:2016bsd, Ge:2016zro, Contino:2016jqw, Biekotter:2016ecg, deBlas:2016ojx, Denizli:2017pyu, Barklow:2017suo, Brivio:2017vri, Barklow:2017awn, Khanpour:2017cfq, Englert:2017aqb, panico,Franceschini:2017xkh, banerjee1, Grojean:2018dqj,Biekotter:2018rhp, Goncalves:2018ptp,Gomez-Ambrosio:2018pnl, Freitas:2019hbk,Banerjee:2019pks, Banerjee:2019twi, Biekotter:2020flu}. As we approach higher integrated luminosities, very precise EFT limits will become achievable. This is, in particular true, because, with a higher luminosity, we will gain the ability to probe the high energy tails of various distributions accurately. This can lead to very precise bounds on SMEFT operators whose contributions grow with energy with respect to the SM.

As far as the Higgs and electroweak physics is concerned refs.~\cite{Franceschini:2017xkh, banerjee1} identified a four-dimensional subspace of the full 59 dimensional space of dimension-6 operators that can be measured very accurately in the diboson processes, $pp \to Vh/VV$, ($V=W^{\pm},Z$) at high energies (see also ref.~\cite{Liu:2018pkg}). That the same set of four operators control both double gauge boson production and Higgs-strahlung is a consequence of the Goldstone Boson Equivalence theorem~\footnote{As a consequence of this theorem, the $Vh/VV$ production amplitudes are equivalent to the amplitude for producing different components of the Higgs doublet in the high energy limit. One can thus connect these amplitudes in the SM as well as the SMEFT using the full $SU(2)_L$ symmetry that is restored at high energies.}. These four directions in the EFT space were dubbed the `high energy primaries' in ref.~\cite{Franceschini:2017xkh}. It was shown that by utilising the quadratic energy growth of the contributions of these operators with respect to the SM, the LHC sensitivity to probe these operators can far surpass LEP bounds.

\begin{figure}[!t]
\centering
\includegraphics[width=0.6 \textwidth]{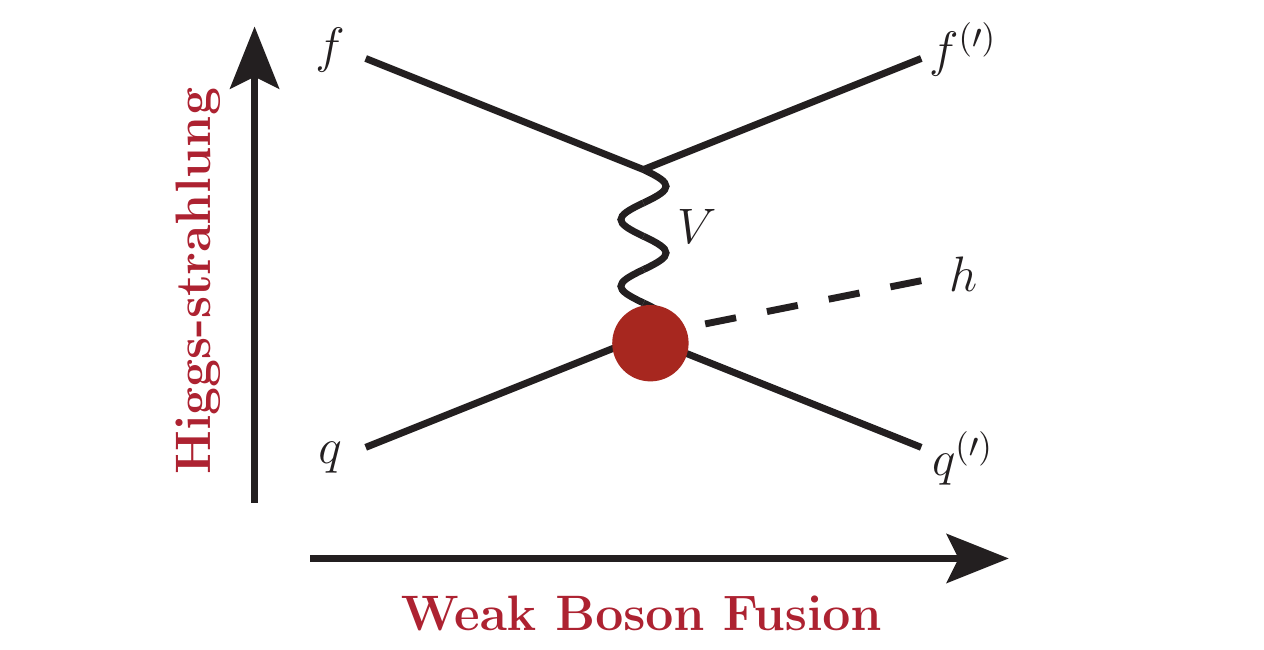}
\caption{Figure shows the crossing symmetry that exists between the Higgs-strahlung and VBF Higgs production processes. The amplitudes for the two processes are the same up to an exchange of the Mandelstam variables, $s \leftrightarrow t$. As a result the same four directions in SMEFT space control VBF Higgs production at high $t$ and Higgs-strahlung at high $s$. The figure has been produced with the help of the {\sc JaxoDraw} package~\cite{Binosi:2008ig}.}\label{fig:vbf-hs}
\end{figure}

In this work, we show that these same high-energy primaries are also sufficient to completely determine the SMEFT amplitude for Higgs production in the Vector Boson Fusion (VBF) channel if the transverse momentum of the Higgs boson is large. The reason for this is a crossing-symmetry that exists between the VBF and Higgs-strahlung diagrams, as shown in Fig.~\ref{fig:vbf-hs}. This implies that the two processes have the same amplitude up to an interchange in the Mandelstam variables, $s\leftrightarrow t$. Thus VBF Higgs production probes the same four operators at large $t$ as Higgs-strahlung at large $s$. Furthermore, one can also extend the equivalence theorem argument used in the diboson case in ref.~\cite{Franceschini:2017xkh} to this case and connect the VBF production of Higgs and gauge bosons.

Thus, the processes, $pp\to VV, Vh$ and VBF production of Higgs or gauge bosons, which are entirely different from each other from a collider physics point of view, actually probe, in a very precise manner, the same set of four operators at high energies. Combining these processes can thus give us the best bounds on the high-energy primaries. Apart from the apparent statistical advantage, it is crucial to combine all these processes because each of them probes a unique linear combination of the four operators; all these processes should, thus, be included to eliminate all flat directions.  

As one of the important results of this work, we will present the linear combination of the four operators that are probed by VBF Higgs production. In this work, we carry out a thorough collider analysis of the $h\to \gamma \gamma$ channel and the three final states from the  $h\to \tau^+ \tau^-$ channel, namely, the hadronic, semi-leptonic and fully leptonic final states. We find that including all these channels is important as their sensitivity to the EFT effects is comparable. In the end, we obtain projections, much stronger than LEP bounds, on the high-energy primaries.

\begin{figure}[!t]
\centering
\includegraphics[width=0.9 \textwidth]{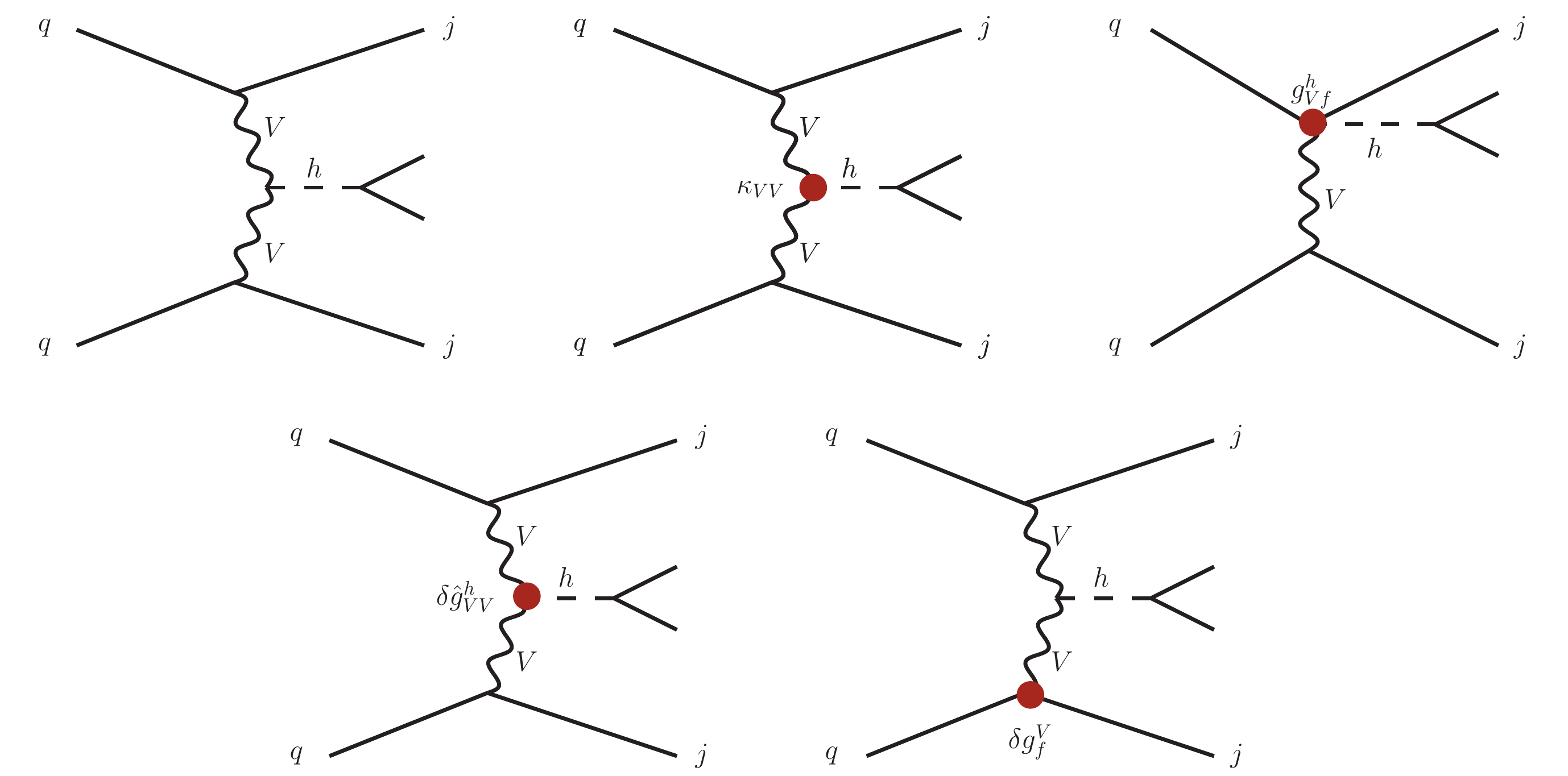}
\caption{Figure shows representative VBF topologies in the SM and in the SMEFT. The red dots signify modified vertices from the EFT couplings in \eq{lagr}. The figure has been produced with the help of the {\sc JaxoDraw} package~\cite{Binosi:2008ig}.}\label{fig:vbf-eft}
\end{figure}

\section{VBF Higgs production at high transverse momentum in the D6 SMEFT}
\label{sec:theory}
The vertices in the dimension-6 (D6) lagrangian that contribute to the VBF Higgs production are the following,
\bea
\Delta {\cal L}_6&\supset& \sum_f {\delta g^Z_{f}} Z_\mu \bar{f} \gamma^\mu f +\delta g^W_{ud} (W^+_\mu \bar{u}_L \gamma^\mu d_L+h.c.) \nonumber \\
&+& \delta{\hat{g}^h_{WW}}\, \frac{2 m_W^2}{v} h W^{+\, \mu} W^-_\mu+ \delta  \hat{g}^h_{ZZ}  \frac{2 m_Z^2}{v}\, h \frac{Z^\mu Z_\mu}{2} \nonumber \\ 
&+& \sum_f g^h_{Zf}\,\frac{h}{v}Z_\mu \bar{f} \gamma^\mu f+g^h_{Wud}\,\frac{h}{v}(W^+_\mu \bar{u}_L \gamma^\mu d_L+h.c.)\\
&+&{\kappa_{\gamma \gamma }}\frac{h}{v} A^{\mu\nu} A_{\mu\nu+}\frac{h}{v} \kappa_{WW}W^{+\, \mu\nu}W^-_{\mu\nu}
+\kappa_{ZZ}\,\frac{h}{2v} Z^{\mu\nu}Z_{\mu\nu}+\delta \hat{g}^h_{\tau\tau}\frac{\sqrt{2} m_\tau}{v} h \tau^+ \tau^- \nonumber \; ,
\label{lagr}
\eea
where we have expanded the D6 SMEFT Lagrangian to obtain lower dimension terms in the broken phase, taking $\alpha_{em}$, $m_Z$ and $m_W$ as the input parameters. Any correction to the SM vector propagators, such as $V_\mu V^\mu, V_{\mu\nu} V^{\mu\nu}$ and $V_{\mu\nu} F^{\mu\nu}$, have been eliminated in favor of the vertex corrections following  refs.~\cite{Gupta:2014rxa, Pomarol:2014dya}. The ${\delta g^Z_{f}}$ and $g^h_{Zf}$ couplings include only a single generation of fermions such that $f=u_L, d_L, u_R \; \textrm{and} \; d_R$. However, we will assume that these couplings are extended to all generations in a flavour universal way which is well justified if we assume Minimal Flavour Violation (MFV)~\cite{DAmbrosio:2002vsn}.

\begin{table}[t]
\begin{center}
\small
\begin{tabular}{l|l}
SILH Basis&Warsaw Basis\\\hline\hline
\rule[-1.2em]{0pt}{3em}$\displaystyle{\cal O}_W=\frac{i}{2}\left( H^\dagger  \tau^a \lra {D^\mu} H \right )D^\nu  W_{\mu \nu}^a$&$\displaystyle{\cal O}^{(3)}_Q= (\bar{Q}\sigma^a\gamma^\mu Q)(i H^\dagger \sigma^a\lra D_\mu   H)$\\
\rule[-1.2em]{0pt}{3em}$\displaystyle{\cal O}_B=\left( H^\dagger  \lra {D^\mu} H \right )\partial^\nu  B_{\mu \nu}$&$\displaystyle{\cal O}_Q=  (\bar{Q} \gamma^\mu Q)(i H^\dagger \lra D_\mu   H)$\\
\rule[-.6em]{0pt}{1.5em}$\displaystyle{\cal O}_{HW}=i g(D^\mu H)^\dagger\sigma^a(D^\nu H)W^a_{\mu\nu}$&
$\displaystyle{\cal O}^u_R= (\bar{u}_R \gamma^\mu u_R)(i H^\dagger \lra D_\mu   H)$\\
\rule[-.6em]{0pt}{1.5em}$\displaystyle{\cal O}_{HB}=i g'(D^\mu H)^\dagger(D^\nu H)B_{\mu\nu}$&$\displaystyle{\cal O}^d_R=  (\bar{d}_R \gamma^\mu d_R)(i H^\dagger \lra D_\mu   H)$\\
\rule[-1.2em]{0pt}{3em}$\displaystyle{\cal O}_{2W}=-\frac{1}{2}  ( D^\mu  W_{\mu \nu}^a)^2$&\\
\rule{0pt}{1.5em}$\displaystyle{\cal O}_{2B}=-\frac{1}{2}( \partial^\mu  B_{\mu \nu})^2$&\\
 \end{tabular}
  \caption{Dimension-six operators contributing to VBF Higgs production at high $p^h_T$.}
\end{center}
\label{operators}
\end{table}

We show how these vertices give corrections to VBF Higgs production in Fig.~\ref{fig:vbf-eft}. It is the subprocess, $q V \to q h$, a common part of all these diagrams, that receives corrections from the D6 lagrangian. As this hard process is a $2 \to 2$ process, its amplitude can be completely specified by two variables, for example, the Mandelstam variable, $t$, and an angle. Up to the leading terms in $t/m_Z^2$ in the EFT correction, we obtain for the $\mathcal{M}(qV_{T,L}\to q h)$ amplitude,
\begin{equation}
  \label{amps}
\begin{split}
q Z_T \to q h&: \, {g^Z_{{f}}} \frac{ \epsilon^*\cdot J_{f}}{v}  \frac{2  m_Z^2}{t}\  \,  \, \Bigg[1+ \left(\frac{g^h_{Zf}}{g^Z_f}+  \kappa_{ZZ}\right)  \frac{t}{2 m_Z^2}  \Bigg] \, ,\\
q Z_L \to q h&: \, {g^Z_{{f}}} \frac{ q\cdot J_{f}}{v}  \frac{2  m_Z}{t}\  \,  \, \Bigg[1+ \frac{g^h_{Zf}}{g^Z_{f}} \frac{t}{2 m_Z^2} \Bigg] \,,\\
q W_T \to q h  &: \, {g^W_{{f}}} \frac{ \epsilon^*\cdot J_{f}}{v}  \frac{2  m_W^2}{t}\  \,  \, \Bigg[1+\left(\frac{g^h_{Wff'}}{g^W_f}+  \kappa_{WW}\right)  \frac{t}{2 m_W^2}  \Bigg] \, ,\\
 q W_L \to q h &: \,  g^W_{f}  \frac{ q\cdot J_{f}}{v}  \frac{2  m_W}{t}\  \,  \, \Bigg[1+\frac{g^h_{Wff'}}{g^W_f} \frac{t}{2 m_W^2} \Bigg] \,,
\end{split}
\end{equation}
where $g^Z_{f}=g(T_3^f-Q_f \stw^2)/\ctw$, and $g^W_f={g}/{\sqrt{2}}$; $J^\mu_f= \bar{f} \gamma^\mu f$ is the fermion current, the subscript $L$ ($T$) denotes the longitudinal (transverse) polarisation of the gauge boson, $q$  denotes its four-momentum and $\epsilon$ the associated polarisation vector. The reason we chose to write the EFT corrections to the amplitude as a function of the Mandelstam variable, $t$, and not $s$, can be understood from \eq{amps}. The EFT corrections are functions of only $t$. The additional angular variable required to specify the scattering kinematics does not appear. This is physically important as it means that the EFT corrections grow with the transverse momentum of the Higgs boson as this kinematic variable is highly correlated with $t$.
 
As we discussed already, the $q V \to q h$ subprocess, is related to the Higgs-strahlung process, $qq \to Vh$, by crossing symmetry, as shown in Fig.~\ref{fig:vbf-hs} such that the expressions in \eq{amps} are identical to the corresponding ones for the $qq \to Vh$ process if we interchange $t \to s$. This is very significant as it implies that VBF Higgs production at high transverse momentum probes the same set of EFT operators as $qq \to Vh$ at high energies. 

\begin{table*}[t]
\begin{center}
\small
\begin{tabular}{c|c}

&EFT directions probed by high energy $ff \to Vh$ production \\\hline\hline
 SILH Lagrangian~\cite{Giudice:2007fh} &$  \frac{g}{c_{\theta_W}}\frac{m_W^2}{\Lambda^2}(2 |T_3^f|\hat{c}_W-2t_{\theta_W}^2 Y_f \hat{c}_B)$\\
 BSM Primaries~\cite{Gupta:2014rxa} &$   \frac{2 g}{c_{\theta_W}}Y_f t_{\theta_W}^2 \delta \kappa_\gamma+2 \delta g^Z_{f}- \frac{2 g}{c_{\theta_W}}(T^f_3 c_{\theta_W}^2 + Y_f s_{\theta_W}^2)\delta g_1^Z $\\
 Universal observables &$ \frac{2 g}{c_{\theta_W}}Y_f t_{\theta_W}^2  (\delta \kappa_\gamma-\hat{S}+Y)- \frac{2 g}{c_{\theta_W}}(T^f_3 c_{\theta_W}^2 + Y_f s_{\theta_W}^2)\delta g_1^Z- \frac{2 g}{c_{\theta_W}}T^f_3 W$\\
 High Energy Primaries~\cite{Franceschini:2017xkh} &$ -\frac{2 m_W^2}{g c_{\theta_W} }(|T_3^f| a_q^{(1)}-T_3^f a_q^{(3)}+(1/2-|T_3^f|)a_f)$\\
 \end{tabular}
  \caption{ The contact interaction couplings $g^h_{Zf}$, where  $f=u_L, d_L, u_R, d_R$ expressed in different EFT parametrisation. For a given $f$ the expression  can  be read off from this table by substituting the corresponding  value of the $SU(2)_L$ and $U(1)_Y$ quantum numbers  $T_3^f$ and $Y_f$. Note that $\hat{c}_W=c_W+c_{HW}-c_{2W}$ and $\hat{c}_B= c_B+c_{HB}-c_{2B}$.}
  \label{dirn}
\end{center}
\end{table*}

For $|t| \gg m_Z^2$, the correction proportional to the contact term coupling, $g^h_{Vf}$ dominates over all other terms\footnote{Note that because of the kinematics, $t$ is always negative in VBF. }. The EFT correction due to $g^h_{Vf}$ grows with $t$ because, unlike the SM diagram, the corresponding diagram in Fig.~\ref{fig:vbf-eft} does not have an intermediate $V$-propagator. The $\kappa_{VV}$ contribution to the transverse amplitude also grows with $t$. This contribution, however, cannot interfere with the dominant longitudinal piece of the SM amplitude and is thus sub-leading with respect to the ${g}^h_{Vf}$ contribution. The  EFT corrections due to the couplings, $\kappa_{\gamma \gamma},\delta g^V_{f}, \delta \hat{g}^h_{\tau\tau}$ and $\delta \hat{g}^h_{VV}$, which include corrections to the Higgs decay,  do not grow with $t$ at all. 

We have checked explicitly that  at high $|t|$ only the ${g}^h_{Vf}$ contributions are important and the effects of the other couplings are negligible provided all the couplings  have a similar size, which is a reasonable assumption if a single cut-off is assumed for the different operators. If this assumption is relaxed, however, it is not immediately clear that only the ${g}^h_{Vf}$ couplings are important at high energies because the different anomalous couplings in Eq.~\ref{lagr} will be constrained at different level at the HL-LHC. We discuss this possibility in Appendix~\ref{subdominant} in detail and show that  the ${g}^h_{Vf}$ contribution dominates even if we take into account the different level of expected constraints on ${g}^h_{Vf}$, $\delta g^V_{f}, \delta \hat{g}^h_{\tau\tau}$ and $\delta \hat{g}^h_{VV}$. Finally in  Appendix~\ref{subdominant} we also show that the process $qq \to h jj$ involving an enhanced $h q \bar{q}$ coupling gives a negligible contribution in our analysis framework once constraints from other processes on $h q \bar{q}$ are taken into account.

Thus, VBF Higgs production at high transverse momentum is controlled by the five contact interaction couplings:  $g^h_{Zf}$, with $f=u_L, u_R,d_L$ and $d_R$ and $g^h_{Wud}$. The operators contributing to these five couplings in the Warsaw basis are shown in Table~\ref{operators}. These contribute to these five anomalous couplings as follows, 

\bea
g^h_{Zu_L}&=&- \frac{g}{c_{\theta_W}}\frac{v^2}{\Lambda^2}(c^{1}_Q-c^{(3)}_Q)\nn\\
g^h_{Zd_L}&=&- \frac{g}{c_{\theta_W}}\frac{v^2}{\Lambda^2}(c^{1}_Q+c^{(3)}_Q)\nn\\
g^h_{Zu_R}&=&- \frac{g}{c_{\theta_W}}\frac{v^2}{\Lambda^2}c_{uR}\nn\\
g^h_{Zd_R}&=&- \frac{g}{c_{\theta_W}}\frac{v^2}{\Lambda^2}c_{dR}.
\label{warsaw}
\eea
The coupling $g^h_{Wud}$ is actually not independent of the above four contact interactions at the dimension-6 level, and is given by,
\bea
g^h_{Wud}=\ctw\frac{g^h_{Zu_L}-g^h_{Zd_L}}{\sqrt{2}}.
\eea
Thus, only the four $g^h_{Zf}$ couplings are independent and these completely determine the EFT deviations for the VBF Higgs production at high transverse momentum.

In Table~\ref{dirn}, we also show the mapping of these four $g^h_{Zf}$ couplings to other EFT parametrisations. In the first row of Table~\ref{dirn}, we present the contributions of the universal (bosonic) operators of the SILH Lagrangian. We then show how these four couplings can be predicted/constrained by other independent measurements. The second row provides the mapping to the so-called BSM Primary basis of ref.~\cite{Gupta:2014rxa}. In this basis, the correlations between different pseudo-observables are made explicit. For instance, in our case we can see how these 4 Higgs anomalous couplings can be predicted in terms of other measurements, namely, the couplings $\delta g^Z_{f}$ defined in \eq{lagr} that are strongly constrained by $Z$-pole measurements at LEP, and the anomalous TGCs, $\delta \kappa_\gamma$ and $\delta g_1^Z$ (in the notation of ref.~\cite{Hagiwara:1986vm}) that were constrained by the $WW$ production during LEP2. In the fourth row of Table~\ref{dirn}, we write the 4 couplings in terms of only the ``oblique''/universal pseudo-observables, i.e. the TGCs $\delta \kappa_\gamma$ and $\delta g_1^Z$ and the Peskin-Takeuchi $\hat{S}$-parameter~\cite{Peskin:1991sw} in the normalisation of ref.~\cite{Barbieri:2004qk}. For a definition of these observables we refer to the Lagrangian presented in ref.~\cite{Elias-Miro:2013eta} (see also ref.~\cite{Wells:2015uba}). Finally, in the last line of Table~\ref{dirn}, we connect these contact terms to the original definition of the high energy primaries in ref.~\cite{Franceschini:2017xkh}.

\section{Collider Analyses}

In this section, we will provide all the details for our collider studies of the three $h \to \tau^+\tau^-$ channels and the $h \to \gamma \gamma$ channel. Utilising the fact that the EFT and SM contributions have the same form apart from a growth in the Mandelstam variable, $t$, we will use a two-step procedure to isolate our EFT signal. First, we will use sophisticated Neural Network (NN) techniques to optimally discriminate between the SM contribution from the other backgrounds in this section. We will then use the $p^h_T$ distribution to isolate the EFT effects from the SM contribution in the next section.

\subsection{The $h \to \tau^+\tau^-$ channels}
\label{sec:ditau}
The SM Higgs decays 6.27\% of the times into a pair of $\tau$-leptons. However, even though this is a significantly large branching ratio, the $\tau$-leptons are not stable, and hence we obtain three distinct final states, depending on the decay modes of the $\tau$s. The cleanest of these final states comprises two light leptons ($e, \mu$). Thus, we categorise our final states as $\tau_{\ell} \tau_{\ell}$, $\tau_{\ell} \tau_h$ and $\tau_h \tau_h$, where $\tau_h$ is the hadronic remnant of the $\tau$ and is identified as a $\tau$-jet. All of these final states are associated with missing transverse energy, $\slashed{E}_T$, and at least two hard jets. We consider all three possibilities here. We closely follow the ATLAS analysis~\cite{Aaboud:2018pen} and then use multivariate methods to optimally isolate the SM VBF Higgs production from the rest of the backgrounds. Our analysis is done at the centre-of-mass energy of 14 TeV. 

The electron (muon) candidates are required to have minimum transverse momentum, $p_T$, of 15 GeV (10 GeV). The electrons (muons) are further required to be in an absolute pseudorapidity region of $|\eta| < 2.47 \; (2.50)$. Furthermore, the electrons are disallowed in the transition region between the barrel and the endcap ($1.37 < |\eta| < 1.52$). Jets are reconstructed using the anti-$k_t$ algorithm~\cite{Cacciari:2008gp} with a jet parameter of $R=0.4$ and with a minimum $p_T$ of 20 GeV. The maximum allowed pseudorapidity range for the jets is required to be 4.5. In order to reconstruct $b$-jets, jets are matched with B-hadrons within $ \Delta R(B,j)<0.2 $ and b-jets are required to have $|\eta| < 2.5$. We require a flat $b$-tagging efficiency of 70\%. In our setup, a light jet (including $c$-jets) can fake a $b$-jet with a fake-tagging efficiency of 1\%. We tag the hadronic $\tau$s with a tagging efficiency of 65\%. Light jets can fake $\tau$-jets with a probability of 2.5\%.

There are multiple backgrounds to consider for the $\tau\tau$ category. The dominant background comes from $\tau^+\tau^-$ jets excluding the Higgs diagrams. We generate this background keeping in mind that the $\tau$s can also emanate from off-shell photons. We also separately generate $\ell^+\ell^-+$ jets, where $\ell = e, \mu$. The other backgrounds include $t\bar{t}$ which we generate separately for the fully leptonic, semi-leptonic and fully hadronic cases, single top ($tq, tW$ and $t\bar{b}$), $\ell (\tau) \nu+$ jets and $h+$ jets with $h \to WW^{*}$, where the $W$s decay either leptonically or hadronically. The $h+$ jets samples are generated for the SM scenario as well as with the EFT couplings turned on. The Feynman rules are generated using the {\sc FeynRules} package~\cite{Alloul:2013bka}, through which we obtain the {\sc UFO}~\cite{Degrande:2011ua} model. All the samples are then generated within the \madgraph\ version 2.6.5~\cite{Alwall:2014hca} framework. The fragmentation, showering and hadronisation are done using \pythia\ version 8.2~\cite{Sjostrand:2014zea}. For the full setup, we use the LO set of NNPDF2.3 parton distribution function~\cite{Ball:2014uwa} within the {\sc LHAPDF} package~\cite{Buckley:2014ana}. For almost all the samples, we use the following cuts at the generation level: $p_{T,j/b(l)} > 20 \; (8)$ GeV, $|\eta_{j(l)}| < 5 \; (3)$, $\Delta R_{jj/bb/bj/ll/jl/bl} > 0.3/0.3/0.3/0/0.2/0.2$, $m_{j_1j_2} > 480$ GeV, where $j_1 \; (j_2)$ is the hardest (second-hardest) quark in $p_T$ and can be a $b$-quark as well, and $\Delta \eta_{j_1j_2} > 2.5$. The $m_{j_1j_2}$ and $\Delta \eta_{j_1j_2}$ cuts are not applied to the single top samples at the generation level. All our event generations are at leading order (LO) in perturbation theory and we consider flat $K$-factors to roughly emulate the next-to-leading order (NLO) QCD effects. For the weak-boson fusion samples, the $K$-factor is almost a constant at 1.1 as a function of $p_{T,j_1}$~\cite{Greljo:2017spw}. For the $l^+l^-+$ jets, $l\nu+$ jets ($l = e,\mu,\tau$), the NLO QCD $K$-factor is roughly 1 as a function of $p_{T,V}$~\cite{Kallweit:2015dum}, with $V$ being the vector boson $W/Z$. For the $t\bar{t}$ samples, we estimated the NNLO $K$-factor be around 1.63~\cite{twiki}. For the single top channel, there are three sub-processes, \textit{i.e.}, $t$-channel, $s$-channel and associated $Wt$ production. The most dominant of these three sub-processes is the $q b \to q' t$ channel followed by the $b g \to tW$ channel. The smallest contribution comes from $q \bar{q}' \to t \bar{b}$. Upon following ref.~\cite{Kant:2014oha}, we consider a conservative $K$-factor of 1.1~\footnote{The NLO electroweak (EW) corrections have not been considered in this paper. As can  be seen from ref.~\cite{Kallweit:2015dum}, the NLO QCD+EW $K$-factors for the aforementioned backgrounds can be less than 1 for higher values of $p_{T,V}$. Hence, we are overestimating these backgrounds to some extent.}. 

To validate our analysis, we reproduce the rectangular cut-based analysis in the ATLAS paper~\cite{Aaboud:2018pen} and find very similar results. The details of our rectangular cut-based analyses are mentioned in Appendix~\ref{app:RCBAtata}.

To obtain  our final results we use a Neural Network (NN) analysis. First,  in addition to  the $p_T$ and $|\eta|$ requirements, we also impose the following cuts, i.e. $m_{j_{1,2}} > 500$ GeV, $\Delta \eta_{j_{1,2}} > 2.5$ and $m_{\tau\tau}^{\textrm{col.}} < 300$ GeV. $m_{\tau\tau}^{\textrm{col.}}$ is the di-tau collinear mass~\cite{Elagin:2010aw}. The variables used in the NN training are shown in Table~\ref{tab:train_params} in Appendix~\ref{app:NN}. In order to prevent the NN from concentrating on the $m_{\tau\tau}$ peak, the sensitivity on the observable has been limited to 5 GeV bins. Table~\ref{tab:NNresults} shows the neural network results for the SM $h\to \tau \tau$ events as well as the other backgrounds,  divided into three sub-regions, namely hadronic, semileptonic and leptonic. The first row shows the number of events at 0.3 ab$^{-1} $ luminosity after the preprocessing mentioned before and the following row shows the yielding number of events after the classification. The procedure and detailed results regarding the neural network are discussed in Appendix~\ref{app:NN}.

\subsection{The $h\to \gamma \gamma$ channel}
\label{sec:diphoton}
Although the diphoton channel suffers from low branching fractions, due to its clean topology, it is relatively easy to separate from the background. With this in mind, we consider diphoton production with two jets topology to single out the VBF channel to achieve higher sensitivity in the aforementioned EFT operators further. By loosely following ref.~\cite{Aaboud:2018xdt}, we construct two workspaces where first we design a cut-and-count based analysis. Then we studied on a Neural Network (NN) architecture which observed to increase our sensitivity. 

Although Higgs-less diphoton with multijet production is the primary background in this channel, it has been shown that in low energy regimes, fake photons can have a significant impact on certain signal regions. The overall fractions of the background sources are presented as $78.7\%$ from Higgsless diphoton channels, $18.6\%$ from single-photon channels and $2.6\%$ from multijet channels~\cite{Aaboud:2018xdt}. It is important to note that these fractions drastically change depending on the phase-space and the efficiency of the jet vertex tagging algorithm~\cite{ATLAS-CONF-2014-018}, where it has been shown that such techniques can reduce fake photon rates below $0.3\%$ especially at higher energies~\cite{ATLAS-CONF-2014-018, Aad:2013aa, PhysRevD.85.012003}. To test this hypothesis, we generate the SM and other background samples using the aforementioned framework. All samples are generated with a specific set of cuts at the matrix-element level; minimum jet $p_T$ is taken to be 30 GeV, two leading jets' invariant mass is chosen to be greater than 500 GeV, and the pseudorapidity separation between the two leading jets is required to be greater than 1.5. As presented in Appendix~\ref{sec:RCBAgaga}, these set of cuts has been chosen with respect to our cut-flow to populate the phase-space that is crucial for this analysis. The generated events are further showered and hadronised via \pythia\ version 8.2~\cite{Sjostrand:2014zea}.

The analysis of the event samples is performed within \madanalysis\ version 1.8~\cite{Conte:2018vmg}. The hadronised events are reconstructed using \fastjet\ version 3.3.2~\cite{Cacciari:2011ma} with the anti-k$_T$ algorithm~\cite{Cacciari:2008gp}, where the radius parameter has been chosen to be 0.4 with minimum transverse momentum of a reconstructed jet at 30 GeV. In order to simulate a simple detector environment, we apply particular tagging efficiencies on the $b$-jets, $c$-jets, hadronic taus and light jets. The tagging criteria are the same as discussed in the di-tau subsection~\ref{sec:ditau}.

In order to get definitive objects, detailed preselection requirements are applied. A photon candidate is required to have a minimum 25 GeV transverse momentum and is chosen to be within $ |\eta| < 2.37 $ and all the photon candidates are required to be separated from each other with $ \Delta R > 0.4 $. On the other hand, a jet candidate is required to be within $|\eta|<4.5$. A clear distinction between photon and jet objects is essential in this analysis in order to suppress the background that might arise from misidentified objects. For this reason, we require the two photons which have a maximum of 15\% hadronic activity within a cone radius of 0.4. After this point we branched our framework into two where cut-based analysis has been discussed in Appendix~\ref{sec:RCBAgaga} and NN analysis has been discussed in Appendix~\ref{app:NN}. Table~\ref{tab:NNresults_sb} shows the NN results presented at 3 ab$ ^{-1} $ integrated luminosity. It shows the event yield for the SM Higgs contribution as well as the other backgrounds at preprocessing stage and for the classifier output for this channel.
\begin{table}
\centering
	\begin{tabular}{l|c|c|c|c}
		& \multicolumn{2}{c|}{Diphoton} &  \multicolumn{2}{c}{Ditau Hadronic} \\\hline\hline
		& Other Background & SM Higgs & Other Background & SM Higgs  \\\hline
		Preprocessing &  11710  & 4621 & 493756 & 27042  \\
		Classifier output & 2251  & 3677 & 69561  & 21897  \\
		\multicolumn{5}{c}{}\\
		&  \multicolumn{2}{c|}{Ditau Semileptonic} &  \multicolumn{2}{c}{Ditau Leptonic} \\\hline\hline
		& Other Background & SM Higgs& Other Background & SM Higgs \\\hline
		Preprocessing & 4190714 & 32343 & 9191181  & 9401 \\
		Classifier output & 91803 & 21469 & 14408  & 3503 \\
	\end{tabular}
	\caption{Table shows NN results  at 3 ab$ ^{-1} $ where we present the number of events for SM VBF Higgs production and the rest of the background before and after NN classification.\label{tab:NNresults_sb}}
\end{table}

\begin{figure}[!t]
	\centering
	\includegraphics[scale=0.5]{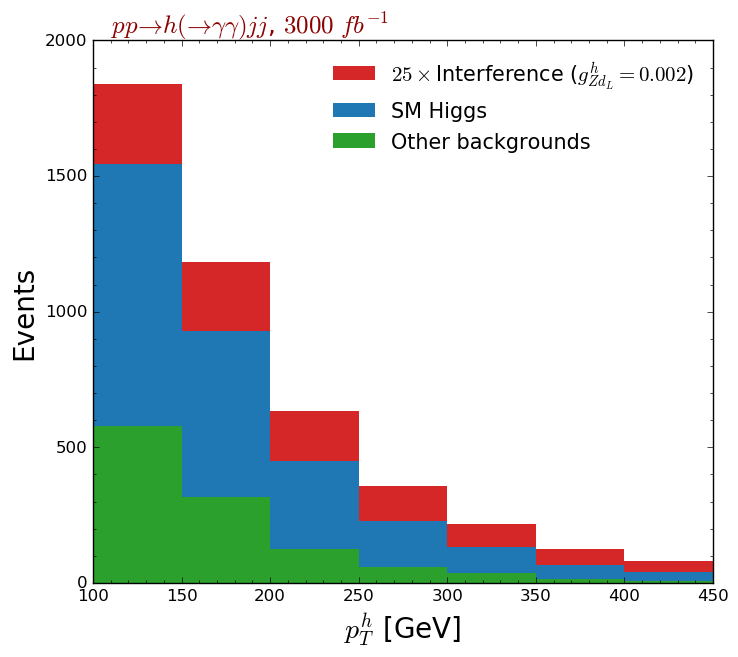}
	\caption{The final distribution with respect to the transverse momentum of the Higgs boson, $p^h_T$, for all the events in the diphoton channel that survive NN classification. The red represents the  EFT interference contribution; blue represents the SM VBF Higgs contribution and green shows the other background sources which are all presented with 3 ab$^{-1}$ data. The  EFT contribution corresponds to interference coming from $g^h_{Zd_L}=0.002$ amplified 25 folds and with vanishing values for all other anomalous couplings in \eq{lagr}.}\label{fig:histo}
\end{figure}
\section{Projected sensitivity for  EFT couplings}

In this section, we present the final sensitivity projections for the EFT couplings. The NN techniques used to optimally isolate the SM VBF Higgs contribution from the other background processes in the previous section also isolate our signal, the EFT interference contribution. This is because as shown in Sec.~\ref{sec:theory} the dominant EFT contributions have a matrix element that is the same as the SM apart from growth with the magnitude of the Mandelstam variable $|t|$. We will now use this growth with $|t|$  to distinguish the EFT interference contribution from the SM; we will utilise the distribution of events with respect to $p^h_T$, a variable that is highly correlated to $t$, as the discriminant. We show the $p^h_{T}$ distribution for the diphoton channel  in Fig.~\ref{fig:histo}. The EFT interference contribution can be seen to grow as a fraction of the SM contribution with $p^h_T$.  To derive the projected sensitivity for the EFT couplings, we define a  $\chi^2$ function as follows.
\begin{eqnarray}
\chi^2 = \sum_{i}^{N}\frac{(N^{exp}_i - N^{obs}_i)^2}{\sigma_i^2}\ ,\nonumber
\end{eqnarray}
 where we take the SM as our null hypothesis. $N^{exp}_i$, denotes the expected number of events in the SM  for the $i$th bin in the $p^h_T$ distribution.  We will then assume that the number of events observed in the $i$-th bin, $N^{obs}_i$ is different from the SM due to the presence of EFT couplings. Finally, $\sigma_i$ includes both the statistical and systematic uncertainties, 
\bea
\sigma_i=\sqrt{N^{exp}_i +(\Delta_{sys} N^{exp}_i)^2}\ ,\nonumber
\eea
 $\Delta_{sys} $ being the percentage of systematic uncertainty. 
 
 As can be seen from Fig.~\ref{fig:histo},  although the   EFT interference contribution steadily grows with $p^h_T$ as a fraction of the SM, the absolute value for the excess keeps decreasing. As a result of the $\chi^2$ function initially increases with $p^h_T$, peaks at an intermediate value around $p^h_T\sim 300$ GeV and then decreases again.  As discussed in Sec.~\ref{sec:theory} the four contact couplings $g^h_{Zf}$,  with $f=u_L, d_L, u_R, d_R$, give dominant contributions in the high $p^h_T$ region. In a hadron collider, it is impossible to disentangle initial states for the process $qV_{T,L}\to q h$  that gets corrections from these contact terms (see Fig.~\ref{fig:vbf-eft}). Thus only a linear combination of these four contact couplings appears in the EFT interference term at a given $p^h_T$. As  bins around $p^h_T\sim 300$ GeV  yield maximum sensitivity, the direction probed by VBF Higgs production turns out to be the above  linear combination at this $p^h_{T}$ value,
\bea
g^{h(VBF)}_{Zf}&=& g^h_{Zu_L} -0.94~g^h_{Zd_L}   - 0.15~g^h_{Zu_R} + 0.04~g^h_{Zd_R}\nn\\
&=&(-0.04~c^1_Q +1.4~c^{(3)}_Q +0.1~c_{uR}-0.03~c_{dR})\xi
\label{vbfdir}
\eea 
where $\xi=v^2/\Lambda^2$. Here the second line expresses the EFT direction in terms of Warsaw basis operators defined in Table~\ref{operators}.   If $p^h_T$ is varied,  the coefficient of $g^h_{Zd_L}$ varies only by a few per cent whereas the coefficients of $g^h_{Zu_R}$ and $g^h_{Zd_R}$   can decrease by as much as 30 $\%$. The left-handed couplings dominate the above direction as the $W$-boson luminosity is much larger than the $Z$-luminosity in VBF processes and the right-handed couplings cannot contribute to the  $q W_L \to q h$ process. 

For our final sensitivity estimate we combine all four final states by adding their individual $\chi^2$ functions. We find each final state has a comparable contribution to the final  $\chi^2$, value which emphasises the importance of including all these four channels. Including only the bins for which $p^h_T<400$ GeV we obtain our final bound for an integrated luminosity of 3 ab$^{-1}$ (0.3 ab$^{-1}$) and $ 10\% $ systematic uncertainty as  
\bea
|g^{h(VBF)}_{Zf}|<0.003\ (0.005)
\label{fbound}
\eea
 at $ 68\% $ CL. Assuming the Wilson coefficients in \eq{warsaw} are ${\cal O}(1)$ the above bound can be translated to the following bound on the cut-off scale,
 \bea
 \Lambda>3.9 {\rm~TeV}\ (\Lambda>2.9 {\rm~TeV})\ .\nonumber
 \label{cutoffbound}
 \eea
 
 The  $p^h_T<400$ GeV cut ensures that most events safely respect the EFT validity requirement $\sqrt{t} < \Lambda$. For strongly coupled UV completions, the values of the Wilson coefficients can be much larger than unity giving much larger values for  $\Lambda$ and thus higher allowed $p^h_T$ values. This, however, would not lead to much better bounds as the most sensitive bins are around $p^h_T\sim 300$ GeV. Our results, thus, do not depend too much on whether the UV completion is weakly or strongly coupled.

\paragraph{Combination with diboson channels:} As we discussed in Sec.~\ref{sec:intro}, the diboson channels $pp \to VV/Vh$ at high energies, and the VBF Higgs production process at high $p^h_T$ considered here, probe the same set of four operators. Of these $WZ$ production was studied in ref.~\cite{Franceschini:2017xkh}, $Zh$ production in ref.~\cite{banerjee1,Banerjee:2019pks} and $Wh$  production in ref.~\cite{Banerjee:2019pks}. Compiling the 68 \% CL HL-LHC bounds obtained in these papers with our result in \eq{fbound}, we obtain  in terms of Warsaw basis operators,
\bea
|(-0.04~c^1_Q +1.4~c^{(3)}_Q +0.1~c_{uR} -0.03~c_{dR})\xi| < 0.003~~~~~&&[VBF] \nn\\
|(-0.18~c^1_Q +1.3~c^{(3)}_Q +0.3~c_{uR} -0.1~c_{dR})\xi| <0.0005~~~~~&&[Zh]\nn\\
|c^{(3)}_Q\xi| <0.0004~~~~~&&[Wh]\nn\\
-0.0004<c^{(3)}_Q\xi <0.0003~~~~~&&[WZ]\label{combine}
\eea
 for 3 ab$^{-1}$ integrated luminosity where $\xi=v^2/\Lambda^2$. It is clear that all these different processes constrain a different direction in four-dimensional space of high energy primaries. As the $WZ$ and $Wh$ process constrain the same direction, the above bounds still leave a flat direction unconstrained.  An additional bound from the $WW$ production process will thus close all flat directions and allow us to bound all the four operators simultaneously. The $WW$ process was studied in ref.~\cite{Grojean:2018dqj} and it is clear from the results that it puts strong bounds on yet another complementary direction. It is, however, difficult to infer the direction probed by the $pp \to WW$ process from the results of ref.~\cite{Grojean:2018dqj} as in this paper the $WW$ and $WZ$ channels have been presented in a combined way including both the interference and EFT squared contributions. 

As mentioned in Sec.~\ref{sec:intro} the VBF production of gauge bosons will also probe the same four-dimensional space. These channels should thus be added to over-constrain the system and maximally constrain the high energy primaries.

\paragraph{Comparison with LEP bounds:} Using Table~\ref{dirn} we can also write this direction in terms of other  pseudo-observables already constrained by LEP,
\bea 
g^{h(VBF)}_{Zf}&=&2~\delta g_{Zu_L} -1.9~\delta g_{Zd_L}   - 0.3~\delta g_{Zu_R} + 0.08~\delta g_{Zd_R}-0.04~\delta \kappa_\gamma   -1.08~\delta g^Z_1 \nonumber\\
g^{h(VBF)}_{Zf}&=&-0.04~(\delta \kappa_\gamma-\hat{S}+Y) -1.08~\delta g^Z_1  -1.4~W
\label{diru}
\eea
where the first line applies to the general case and the second line to the universal case. The LEP bounds on the above pseudo-observables are given by the second column of Table~\ref{lepb}.  The LEP bound on the full direction is thus given by the largest term in the right hand sides of the above equations which is $g^{h(VBF)}_{Zf} \lesssim 1.08~\delta g^Z_1\lesssim 0.03$ which is almost an order of magnitude weaker than the bound in \eq{fbound}. 

One can also assume that there is no cancellation between the different terms in  eq.~\ref{diru}. This allows us to require that each term in the right-hand side respects the bound in \eq{fbound}.  We then obtain the results in the first column of  Table~\ref{lepb}. We see that with this `no tuning' assumption, relative to LEP bounds, the results of this work can lead to much stronger bounds on TGCs, comparable bounds on deviations of $Z$ coupling to quarks and weaker bounds for the oblique parameters.
\begin{table}[!h]
	\begin{center}
		
		\begin{tabular}{c|c|c}
			&Our Projection &LEP Bound\\\hline\hline
			$\delta g^Z_{u_L}$& $\pm0.001~(\pm0.002)$ &$[-0.0042, -0.001]$\\
			$\delta g^Z_{d_L}$& $\pm0.002~(\pm0.003)$&$[0.0013, 0.0033]$\\
			$\delta g^Z_{u_R}$& $\pm0.01~(\pm0.02)$ &$[-0.0071, -0.0001]$\\
			$\delta g^Z_{d_R}$& $\pm0.04~(\pm0.06)$ &$[0.0108, 0.0212]$\\
			$\delta g^Z_1$&$\pm0.003~(\pm0.005)$ &$[-0.03, 0.013]$\\
			$\delta \kappa_\gamma$&$\pm0.08~(\pm0.12)$&$[-0.063, 0.026]$\\
			$\hat{S}$& $\pm0.07~(\pm0.12)$&$[-0.0003, 0.0011]$\\
			$W$& $\pm0.002~(\pm0.004)$& $[-0.0006, 0.0000]]$\\
			$Y$& $\pm0.07~(\pm0.12)$&$[-0.0003, 0.0003]$\\
		\end{tabular}
		\caption{Table shows the comparison of 68 \% CL bounds extracted from the VBF analyses with the existing LEP bounds. The bounds outside the parentheses are projections  for 3 ab$^{-1}$ data and those inside are for 0.3 ab$^{-1}$ data. To get our projection, we demand that each term in~\eq{diru} respects the bound in~\eq{fbound}. The LEP bounds on the $Z$-boson couplings to quarks, $\delta g^Z_f$, are taken from ref.~\cite{Falkowski:2014tna}, the bounds on the charged TGCs are taken from ref.~\cite{LEP2}, the bound on $\hat{S}$ from ref.~\cite{Baak:2012kk}, and the bounds on the  $W,Y$ observables from ref.~\cite{Barbieri:2004qk}.   \label{lepb}} 
	\end{center}
\end{table}
\section{Conclusions}

 It is increasingly being recognised that the LHC is a precision machine. This is because various examples are beginning to appear where certain operators can be probed very precisely, for instance, by studying the high energy tails of different processes. One of the best examples is that of the high energy primaries, the four operators that dominate the high energy tails in diboson production, including the Higgs-strahlung process. 
 In this work, we highlight how VBF Higgs production probes a linear combination of the same operators given by eq.~\ref{vbfdir}.   Our results are complementary to those obtained in the Higgsstrahlung, and diboson processes as all these processes probe different directions in this four-dimensional space (see \eq{combine}). Our final projection for the HL-LHC bounds on the direction corresponding to VBF Higgs production is per mille level (see \eq{fbound}) which translates to a multi-TeV bound on the new physics scale given in \eq{cutoffbound}. These bounds far surpass the existing  LEP bounds (see discussion below \eq{diru} and Table~\ref{lepb}).

As far as Higgs and electroweak physics is concerned,  the highest energy scales the LHC will probe indirectly might well be via a precise measurement of these high energy primaries\footnote{Another example where LHC can indirectly probe very high scales by studying high energy tails is the Drell-Yan process as shown in ref.~\cite{Farina:2016ws}.}. These may therefore become part of the legacy measurements of LHC.  The VBF Higgs production process studied in this work would be an important and integral part of this program.

\appendix

\section{Rectangular Cut-Based Analyses}\label{app:RCBA}

\subsection{The $h \to \tau^+\tau^-$ channels}\label{app:RCBAtata}
Overall, we follow the relevant cuts listed in tables 3 and 4 of ref.~\cite{Aaboud:2018pen}. For the $\tau_{\ell}\tau_{\ell}$ case, we demand $m_{j_1j_2} > 500$ GeV instead of 800 GeV as mentioned in the paper. We use the tight VBF category for the $\tau_h\tau_h$ case. For our lepton isolation, we require that the hadronic activity around an isolated lepton ($e,\mu$) within a cone of $\Delta R = 0.2$, should not exceed 10\% of its $p_T$. With the rectangular cut-based analyses, we get the following results. Following ATLAS, for the rectangular cut-based analysis, we dissect the $\tau_{\ell}\tau_{\ell}$ scenario into the same flavour and opposite flavour cases. For the same flavour case ($ee$ or $\mu\mu$), the  number of events from the  SM Higgs signal ($S$) and other background ($B$)\footnote{We must note that, here and in what follows,  we are loosely referring to the SM VBF as the `signal' ($S$),  and the  the rest of the samples as `background' ($B$). In our final analyses, the SM VBF is of course part of the background and the EFT contribution is the true signal.} at an integrated luminosity ($\mathcal{L}$) of 300 fb$^{-1}$ are 119 and 953, yielding a significance ($S/\sqrt{S+B}) \sim 3.64$. For the different flavour case, $S \sim 149$, $B \sim 1361$ and significance $\sim 3.84$. For the $\tau_{\ell} \tau_h$ case, $S \sim 436, B \sim 10332$ and significance $\sim 4.20$. Finally, for the $\tau_h\tau_h$ case, we obtain $S \sim 686, B \sim 9647$ and significance $\sim 6.74$.

\subsection{The $h \to \gamma \gamma$ channel}\label{sec:RCBAgaga}
As mentioned in section~\ref{sec:diphoton}, we construct a cut-based analysis by loosely following ref.~\cite{Aaboud:2018xdt}. After the preselections mentioned above, in order to identify the VBF channel, we use standard VBF cuts where the $b$-jets are vetoed, and we require at least two jets where the leading two are separated into two hemispheres with $|\Delta\eta|>3 $. To identify the boosted VBF topology, the general recipe requires an invariant mass cut between two leading jets at the order of 300-400 GeV as applied in ref.~\cite{Aaboud:2018xdt}. However, we observe higher sensitivity to the EFT operators achieved when higher $M_{jj} $ requirement is applied. In addition to the isolation requirement presented in section~\ref{sec:diphoton}, we further demand additional angular requirements between the photons and jets to restrict the phase-space for additional emissions. The minimum angular separation between jets and photons, $\Delta R^{min}_{\gamma j}$ is observed to perform as a tremendous discriminatory tool against background rejection. Limiting $\Delta R^{min}_{\gamma j}>1.5$ is observed to separate the background from the signal events without any loss of the desired phase-space. We also require an azimuthal angle separation between the two-jet and two-photon systems. Although this requirement does not propose relatively active discrimination, it has been shown to be a powerful tool to suppress theoretical uncertainties and veto additional jets in the event sample~\cite{Aaboud:2018xdt}. These series of requirements cause largely boosted samples, where although our signal does not show any particular azimuthal separation preference, we observe that the background is dominated by highly separated jets in azimuthal angle. For this reason, we require angular separation between the two leading jets to be less than 2. Finally, the most effective cut was expectedly the invariant mass of the two-photon system, which is chosen to be within $125\pm 3$ GeV. Also, the reconstructed Higgs rapidity is required to lie between the two jets. Table~\ref{tab:diphoton_cutflow} summarises all the cuts and their relative efficiencies for both the signal and the background samples. At the bottom portion of the table, we show various discriminatory variables to asses the quality of the yield events. All results are presented at 0.3~ab$^{-1}$. It is important to note that we also generate single-photon samples to quantify the effect of fake photon contamination in the sample. For this we use the {\sc SFS} module of \madanalysis~\cite{Araz:2020lnp} to simulate a light jet mis-tag rate of $0.3\%$. However, we observe that out of a million events, we do not have any to pass the Higgs mass requirement. Thus, in order to save valuable computation time, we assume that such effects are insignificant in such boosted phase-space regimes.
\begin{table}[!h]
	\centering
	\begin{tabular}{l|cc|cc}
		& \multicolumn{2}{c|}{Background} & \multicolumn{2}{c}{SM Signal} \\\hline\hline
		& Events  & $ \varepsilon $ [\%]   & Events & $ \varepsilon $ [\%] \\\hline
		Presel.                                                 & 369176.1 & -        & 1365.2  & -      \\
		Njet$\geq2$                                             & 286704.2  & 77.66   & 1144.9  & 83.87 \\
		Bjet veto                                               & 274869.6 & 95.87   & 1108.2  & 96.79 \\
		$|\Delta\eta_{jj}| > 3$                                 & 164813.0  & 59.96   & 838.0 & 75.62 \\
		$\eta_{j_1}\cdot\eta_{j_2} < 0$                              & 161844.1  & 98.20   & 827.2  & 98.71 \\
		$M_{jj}>600$ [GeV]                                      & 93105.4 & 57.53   & 658.6  & 79.62 \\
		$N_\gamma=2$                                            & 20244.1  & 21.74   & 432.8 & 65.72 \\
		$I^{R=0.4}_\gamma< 15\%$                                   & 19876.9 & 98.19   & 431.9 & 99.77 \\
		$\Delta R^{min}_{\gamma j}>1.5$                         & 8379.1 & 42.15   & 382.6  & 88.58 \\
		$|\Delta\Phi_{\gamma\gamma,jj}|>1.5$                    & 7896.7  & 94.24   & 373.4  & 97.62 \\
		$\Delta\Phi_{j_1,j_2}<2$                                & 2393.7 & 30.31   & 227.7 & 60.97 \\
		$122<M_{\gamma\gamma}<128$ [GeV]                        & 88.1 & 3.68    & 226.9  & 99.64 \\
		$y^{min}_{j_{1,2}} < y_h < y^{max}_{j_{1,2}}$           & 78.0  & 88.63   & 223.2 & 98.40 \\\hline\hline
		$S/B$                                                      & \multicolumn{4}{c}{286.06\%            } \\
		$S/(B+S)$                                                & \multicolumn{4}{c}{74.10\%             } \\
		$S/\sqrt{B}$                                          & \multicolumn{4}{c}{25.27        } \\
		$S/\sqrt{S+B}$                                          & \multicolumn{4}{c}{12.86        } \\
	\end{tabular}
	\caption{Cut-flow for diphoton channel presented. Both background and signal samples separated in to two columns where on the left of each one yielding number of events are presented at 0.3 ab$ ^{-1} $ and the right column shows the relative efficiency of each cut. At the bottom part certain ratios are presented. Note that what's referred to as signal ($S$) here is the SM VBF. In the final analysis, the SM VBF is part of the background.}\label{tab:diphoton_cutflow}
\end{table}

\section{Neural Network Analysis}\label{app:NN}
In recent years, the particle physics community has been increasingly adapting to the use of {\it deep neural networks} (DNN) in challenging signal characterisation problems~\cite{Baldi:2014pta,deOliveira:2015xxd,Baldi:2016fzo,Caron:2016hib,Chang:2017kvc,Lin:2018cin,Albertsson:2018maf,Guest:2018yhq,Abdughani:2019wuv,Windischhofer:2019ltt,Amacker:2020bmn}. The \texttt{Keras} library~\cite{chollet2015keras} offers a python-based, flexible framework using feed-forward networks~\cite{HORNIK1991251,1165576,HORNIK1989359} to create mashed layers with connected neurons (nodes). In order to increase our sensitivity to the operators presented above, we design a simple workspace to determine achievable sensitivities with different neural network architectures. We assume to have certain common properties to apply on each architecture. Each architecture is optimised using the \texttt{Adam} algorithm~\cite{Kingma2014AdamAM} and to accommodate multi-class classification, sparse categorical crossentropy loss function is used where the crossentropy is defined as
\begin{eqnarray}
\mathcal{H}(p^{truth},p^{pred}) = -\frac{1}{N}\sum_{i=1}^{N}p^{truth}_i\log p^{pred}_i\ . \nonumber
\end{eqnarray}
Here $p^{truth}$ refers to the vector of the truth values and $p^{pred}$ is the vector of prediction probabilities. We use softmax activation in the output, which is essentially a combination of sigmoid functions for each output class. Furthermore, instead of traditionally used sigmoid activation for each layer, we use rectified linear unit (\texttt{ReLU})~\footnote{See ref.~\cite{Aad:2019yxi} for advantages of the \texttt{ReLU} activation over sigmoid function.}. Each model is initialised with a learning rate of $1\times10^{-5}$, and the learning rate decayed to its half if the loss value of the validation sample does not improve for 20 epochs. Each training runs for 200 epochs with a requirement of at least 0.01 unit improvement on the loss of the validation sample. The samples that do not satisfy this condition for 50 epochs are terminated before the end of the 200 epochs. The class weights are normalised with respect to their occurrences in the training sample in order to compensate for the difference of the population of each class. We investigate four different signal regions, namely diphoton, and ditau decaying hadronically (hereafter hadronic), ditau decaying semileptonically (hereafter semi-leptonic) and ditau decaying leptonically (hereafter leptonic). Each sample has a separate set of backgrounds, and in order to save computation time, we only use the dominant background samples that have the greatest impact in a given signal region.

In order to prevent over-training for each signal region, we use the dropout and kernel-regularisation methods. Each layer is required to have 25\% probability of dropping each node in order to prevent dependency on a given parameter. Additionally, each hidden layer is supported via the $L_2 $ kernel regularisation~\cite{L2Regularization} with a penalty strength of $10^{-2}$. This penalty term is directly reflected on the loss as
\begin{eqnarray}
{\rm Loss} := \mathcal{H} + \lambda \sum^{N}_{i=1}(||\omega_i||^2 + ||b_i||^2)\ ,\nonumber
\end{eqnarray}
where $\omega_i$ is the weight of the node, $b_i$ is the bias and $\lambda$ is the penalty strength. Lower and higher values of the penalty strength are tested as well which lead to the signal regions to over train and, in case of the latter, cumulative accuracy has been observed to drop below 70\% respectively.

Each signal region undergoes a specific preprocessing before training. In addition to aforementioned preselection requirements, diphoton sample is required to have at least two jets, and two isolated photons in the data sample and the invariant mass of the two leading jet is required to be at least 550 GeV. In order to prevent the NN from concentrating only on the diphoton invariant mass, we require it to be within $125\pm3 $ GeV window and remove the $M_{\gamma\gamma}$ from the training parameters. Otherwise, it has been observed that the NN avoids all other variables and concentrate solely on $ M_{\gamma\gamma} $ peak, which has been tested up to 5 GeV resolution for the distribution. Although this reduces the accuracy of the test sample significantly, it is necessary to avoid the NN to concentrate only on the sharp invariant mass peak. In order to remain in the realm of VBF, we also veto all $b$-jets and require $|\Delta\eta_{jj}|>1.5$, without requiring them to be on different hemispheres. Following the same recipe, the ditau samples are preprocessed by requiring $M_{jj}>500$ GeV, $|\Delta\eta_{jj}| > 2.5$ and $M_{\tau\tau} < 300$ GeV. For the leptonic final state, we demand two isolated leptons, for the semi-leptonic final state, one isolated lepton and one hadronic tau. Finally, for the hadronic final state, we require two hadronic taus and veto events with isolated leptons. As before, in order to prevent the NN to concentrate on the invariant mass peak of two taus, we require it to have a resolution of 5 GeV. Table~\ref{tab:train_params} summarises all the parameters that have been used for each region. Here, $p^{jj}_T$ refer to the combined vectorial $p_T$ of the two-hardest jets, $\tau_{1h/2h}$ refer to the visible part of the hardest and the second-hardest $\tau$-lepton, which can be $e, \mu$ or $\tau_h$, $m_{T2}$ is the stransverse mass variable~\cite{Baringer:2011nh, Barr:2013tda}, $\Delta \phi$s are the azimuthal angle separations, $x_{1/2}$ are the visible momentum fractions for the two $\tau$ leptons and $\Delta R = \sqrt{\Delta \eta^2 + \Delta \phi^2}$. All the other variables are self-explanatory. For the semi-leptonic case, we also have an additional variable, the transverse mass, $m_T$.

\begin{table}[!h]
	\centering
	\begin{tabular}{l|p{9cm}}
		Ditau Hadronic & $p^{j_1/j_2/\tau_{1h}/\tau_{2h}}_T$, $p^{jj}_T$, $\frac{p^{j_1}_T}{p^{j_2}_T}$, $\Delta R(j_1,j_2)$, $p^{vis}_T(h)$, $p^{miss}_T$, $M_{T2}$, $M_{jj}$, $|\Delta\eta(j_1,j_2)|$, $x_1$, $x_2$, $\Delta\phi(\tau_{1h}/\tau_{2h},\mathbf{p}^{miss}_T)$, $\Delta\phi(j_1,j_2)$, $M_{\tau\tau}^{\textrm{col.}}$, Jet Multip., $\frac{p^{\tau_{1h}}_T}{p^{\tau_{2h}}_T}$, $\Delta R(\tau_{1h},\tau_{2h})$, $\Delta R(\tau_{1h}/\tau_{2h},j_1/j_2)$, $M(j_1/j_2,\tau_{1h}/\tau_{2h})$, $|\Delta\eta(\tau_{1h},\tau_{2h})|$, $M_{\tau_h\tau_h}$ \\\hline
		Ditau Semileptonic & $p^{j_1/j_2/\ell/\tau_h}_T$, $p^{jj}_T$, $\frac{p^{j_1}_T}{p^{j_2}_T}$, $\Delta R(j_1,j_2)$, $p^{vis}_T(h)$, $p^{miss}_T$, $M_{T2}$, $M_{jj}$, $|\Delta\eta(j_1,j_2)|$, $x_1$, $x_2$, $\Delta\phi(\ell/\tau_h,\mathbf{p}^{miss}_T)$, $\Delta\phi(j_1,j_2)$, $M_{\tau\tau}^{\textrm{col.}}$, Jet Multip., $\frac{p^{\ell}_T}{p^{\tau_h}_T}$, $\Delta R(\ell,\tau_h)$, $\Delta R(\ell,j_1/j_2)$, $\Delta R(j_1/j_2,\tau_h)$, $M(j_1,j_2,\ell,\tau_h)$, $|\Delta\eta(\ell,\tau_h)|$, $M_{\ell \tau_h}$, $M_T$\\\hline
		Ditau Leptonic & $p^{j_1/j_2/\ell_1/\ell_2}_T$, $p^{jj}_T$, $\frac{p^{j_1}_T}{p^{j_2}_T}$, $\Delta R(j_1,j_2)$, $p^{vis}_T(h)$, $p^{miss}_T$, $M_{T2}$, $M_{jj}$, $|\Delta\eta(j_1,j_2)|$, $x_1$, $x_2$, $\Delta\phi(\ell_1/\ell_2,\mathbf{p}^{miss}_T)$, $\Delta\phi(j_1,j_2)$, $M_{\tau\tau}^{\textrm{col.}}$, Jet Multip., $\frac{p^{\ell_1}_T}{p^{\ell_2}_T}$, $\Delta R(\ell_1,\ell_2)$, $\Delta R(j_1/j_2,\ell_1/\ell_2)$, $M(j_1,j_2,\ell_1,\ell_2)$, $|\Delta\eta(\ell_1,\ell_2)|$, $M_{\ell\ell}$\\\hline
		Diphoton & $M_{jj}$, $\eta_{j_1/j_2}$, $p^{h/\gamma_1/\gamma_2/j_1/j_2}_T$, $\Delta\phi(j_1,j_2)$, $\Delta\phi(h,j_1/j_2/\gamma_1/\gamma_2)$, $\Delta R(h,j_1/j_2/\gamma_1/\gamma_2)$, $M_{eff}$, $H_T$, $\Delta R(\gamma_1,\gamma_2)$, $y_{j_1/j_2/h}$, $\Delta R^{min}_{\gamma j}$, $\Delta\Phi(h,jj)$, $p^{hjj}_T$, $|\eta_h-0.5(\eta_{j_1}+\eta_{j_2})|$\\
	\end{tabular}
	\caption{Parameters which are used in corresponding NN training. Observables shown with comma, $ \mathcal{O}^{i,j} $, represents a system of corresponding $ i^{th} $ leading and $ j^{th} $ leading particles and the ones shown with slash, $ \mathcal{O}^{i/j} $, represents the usage of the same observable for both $ i^{th} $ leading and $ j^{th} $ leading reconstructed object separately.\label{tab:train_params}}
\end{table}

\begin{figure}[!h]
	\begin{center}
		\includegraphics[scale=.4]{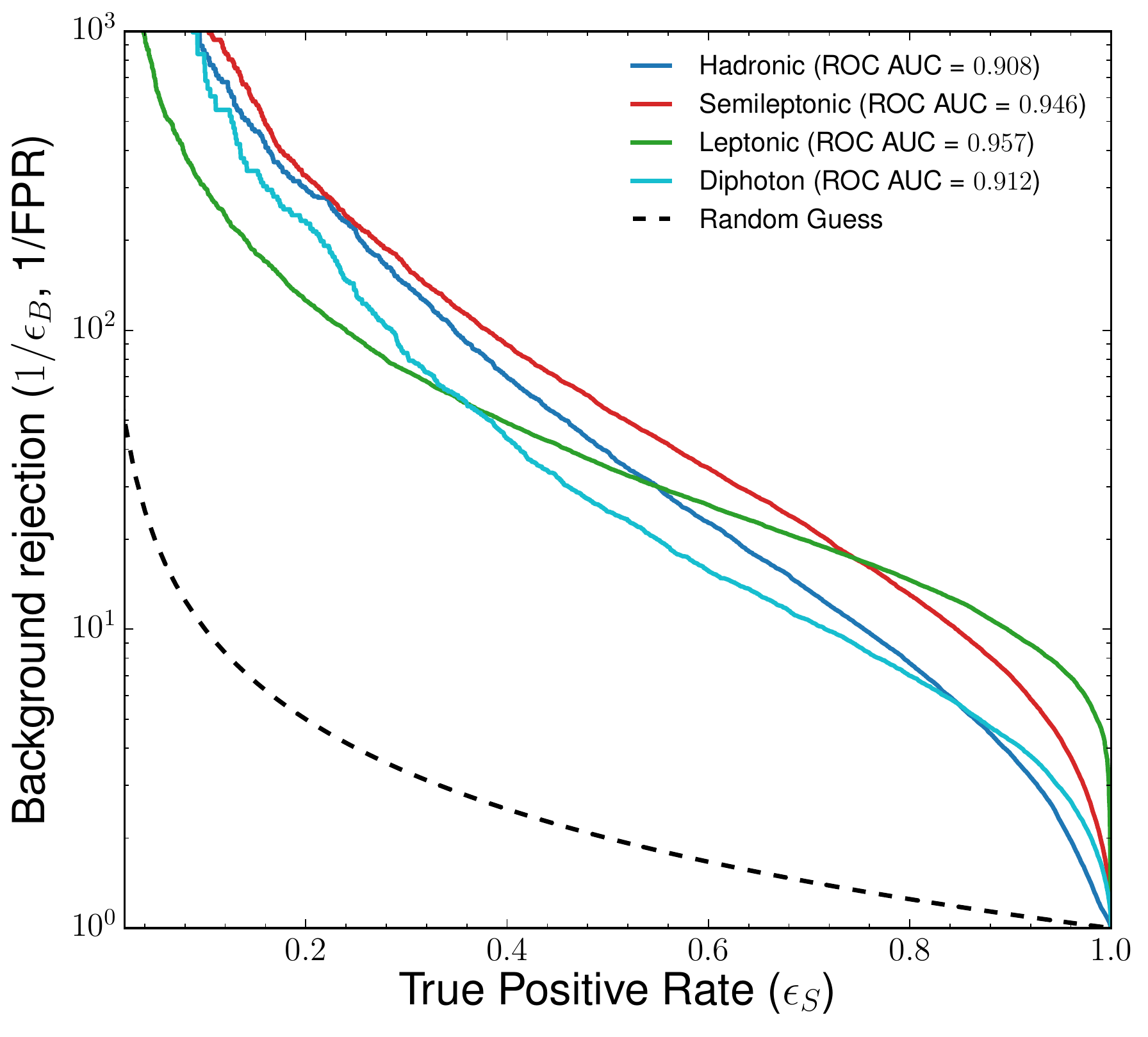} 
	\end{center}
	\caption{ROC curve for diphoton and ditau signal regions. Each label presented with the corresponding area under TPR-FPR curve. The colors blue, red, green and cyan represents ditau hadronic, ditau semileptonic, ditau leptonic and diphoton signal regions.}\label{fig:roc}
\end{figure}

Hyper-parameter optimisation is a challenging problem in machine learning. To further understand the phase-space and the effect of the layers, we devise a simple scanning procedure which starts from a linear model and increases number of hidden layers and nodes depending on the performance of the NN. To simplify the process, the number of nodes is chosen to be a certain multiple of the number of input parameters. Table~\ref{tab:NNresults_sb} shows the general results of the NN where preselection gives the number of events remaining after the preprocessing and the classifier output is the number of events left after the classification process. All the results are presented at 3 ab$ ^{-1}$ integrated luminosity. In Table~\ref{tab:NNresults}, we present corresponding classification results for signal in the training sample. As expected, we observe a larger $S/B$ ratio for diphoton channel with respect to the ditau channels. The test accuracy is measured via 10-fold validation in order to see the fluctuations in the results. Although the diphoton channel gives the least amount of uncertainties, none of the uncertainties goes beyond 4\% of the mean test accuracy. We also present the signal precision, the true positive rate (TPR) and the F1-score~\cite{precision_recall} for the test sample. The last row of the \autoref{tab:NNresults} shows the number of hidden layers and their corresponding number of nodes in each layer. Fig.~\ref{fig:roc} shows the corresponding receiver operating characteristic (ROC) curve where dark blue, red, green and light blue curves represents hadronic, semileptonic, leptonic and diphoton channels. Area under the ROC (AUC) curve with respect to TPR and false positive rate (FPR) has been attached to each label. Black dashed line presents a reference for random guess.
\begin{table}
	\centering
	\begin{tabular}{lcccc}
		& & Ditau & Ditau & Ditau  \\
		& Diphoton &  Hadronic &   Semileptonic &   Leptonic \\\hline\hline
		Test accuracy & $ 80.04\% \pm 0.01\% $ & $ 76.04\% \pm  1.19\%$ & $ 77.23\% \pm3.31\%$ &  $ 70.31\%\pm2.43\% $ \\
		Signal precision &  $ 98\% $ &  $ 92\% $ & $ 75\% $ & $ 94\% $\\
		Signal TPR      &  $ 79\% $ &  $ 81\% $ & $ 37\% $ & $ 66\% $ \\
		Signal F1-Score &  $ 88\% $ &  $ 86\% $ & $ 50\% $ & $ 78\% $\\
		Layers $ \times $ Nodes &  $ 5\times675 $  &  $ 3\times840 $ &  $ 4\times1015 $ & $ 5\times1400 $  \\
	\end{tabular}
	\caption{Results for NN classification presented. Table shows the statistics of the NN calculated for the test sample. Last row of the table shows the number of hidden layers and node per each layer.\label{tab:NNresults} }
\end{table}

\begin{figure}
	\begin{center}
		\includegraphics[scale=.35]{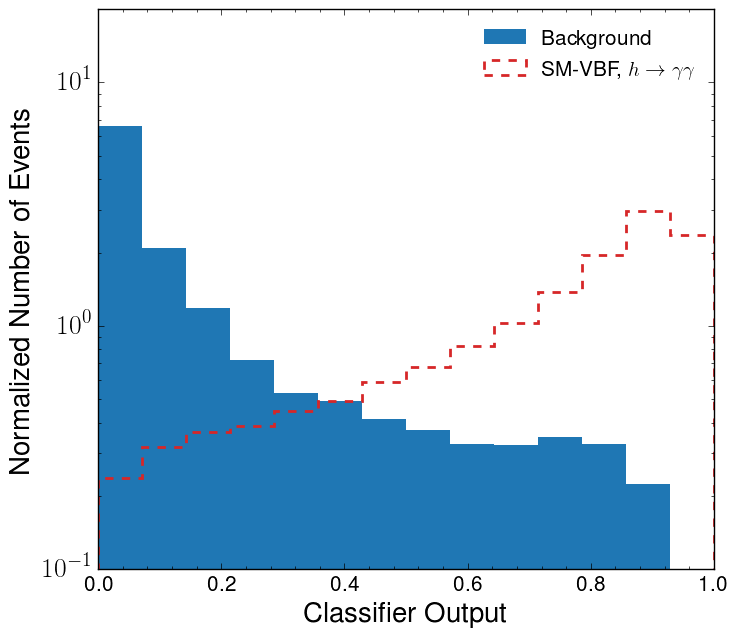}  
		\includegraphics[scale=.4]{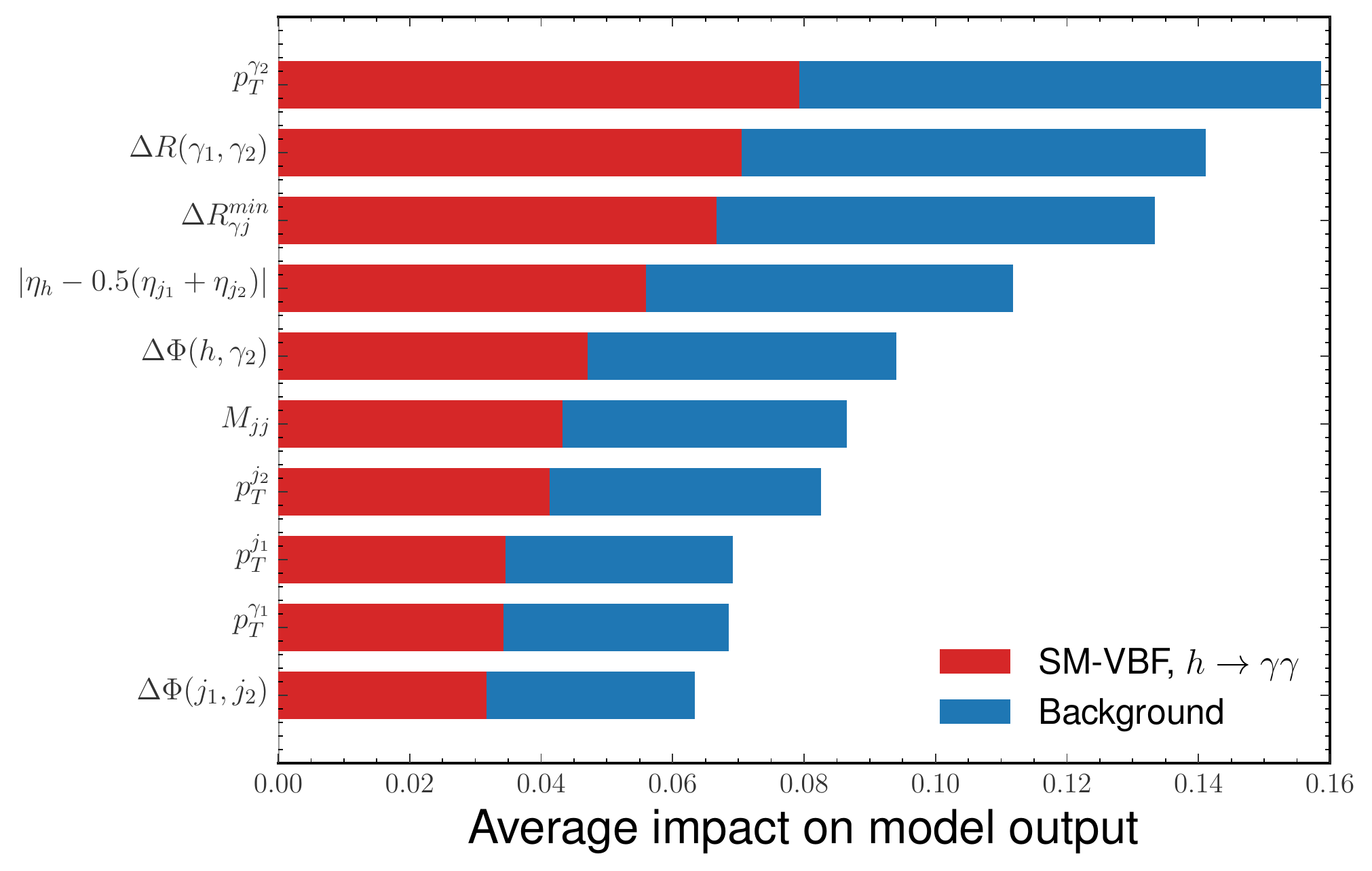} 
	\end{center}
	\caption{Left panel shows the classification output for the diphoton channel and the right panel shows the average {\sc SHAP} value for ten leading parameter that is used in the training. Red shows the signal and blue represents the background in both figures.}\label{fig:diphoton_out}
\end{figure}

\begin{figure}
	\begin{center}
		\includegraphics[scale=.35]{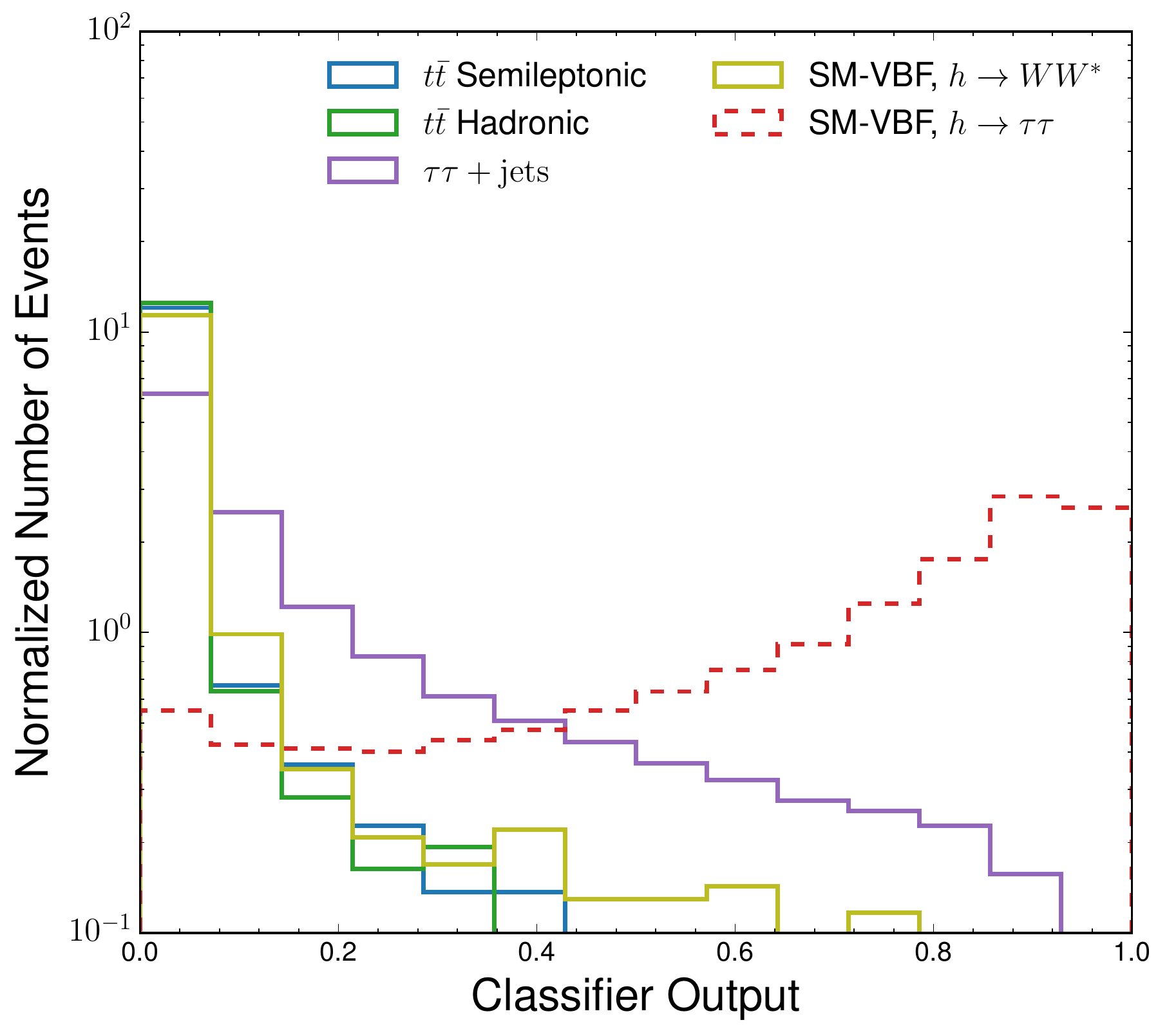}  
		\includegraphics[scale=.4]{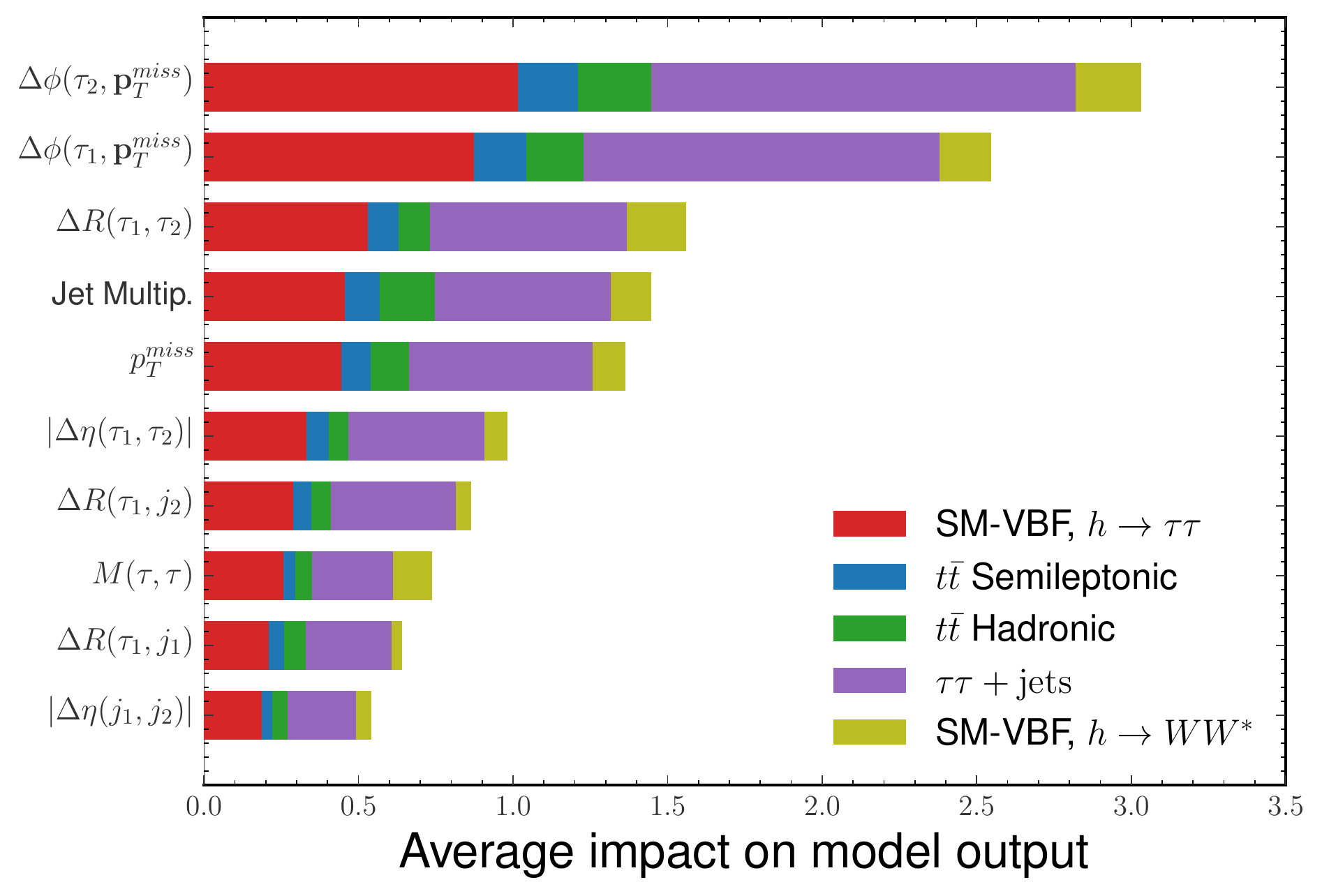} 
	\end{center}
	\caption{Same as Fig.~\ref{fig:diphoton_out} for hadronic channel.}\label{fig:tHtH_out}
\end{figure}

\begin{figure}
	\begin{center}
		\includegraphics[scale=.35]{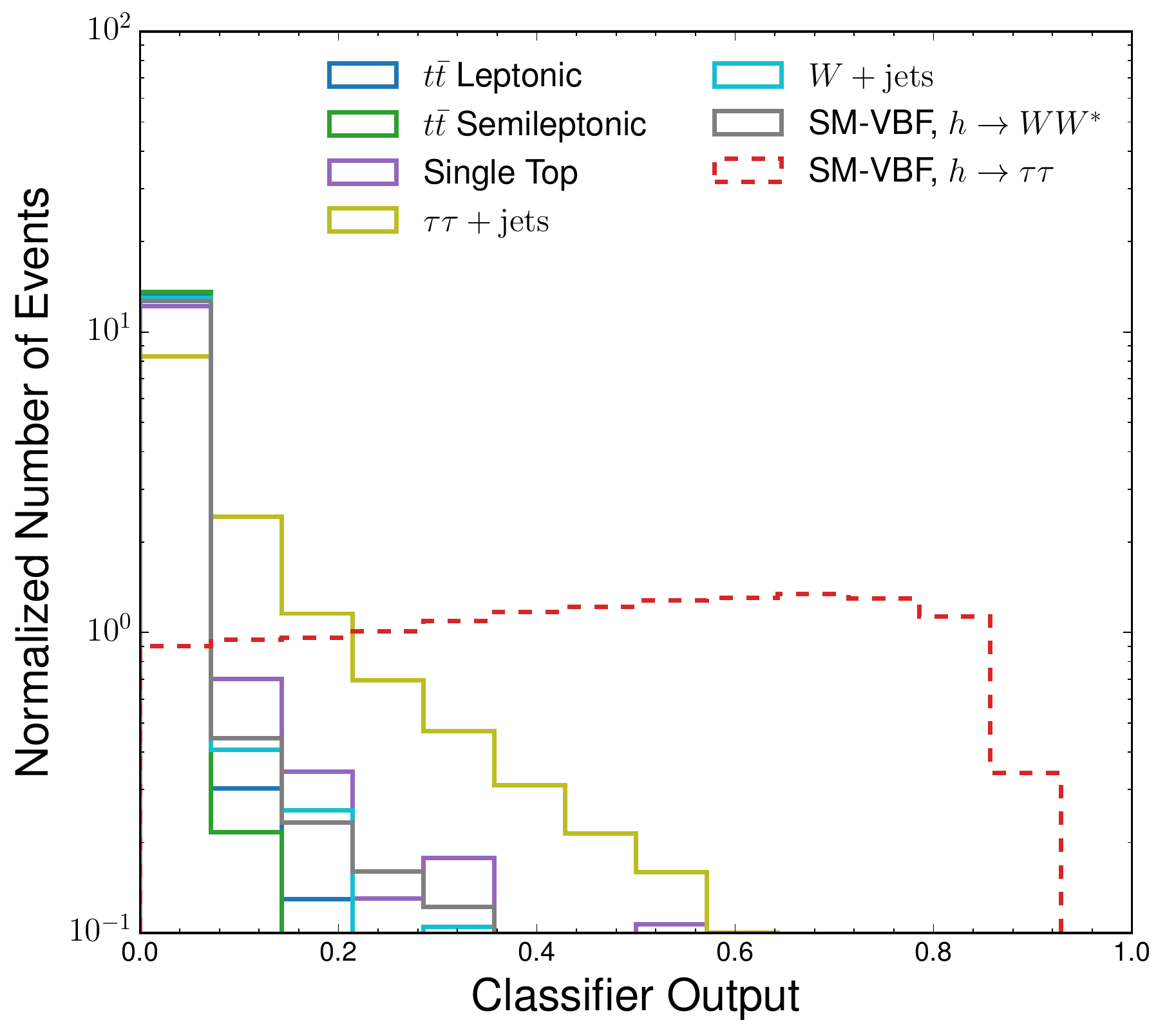}  
		\includegraphics[scale=.4]{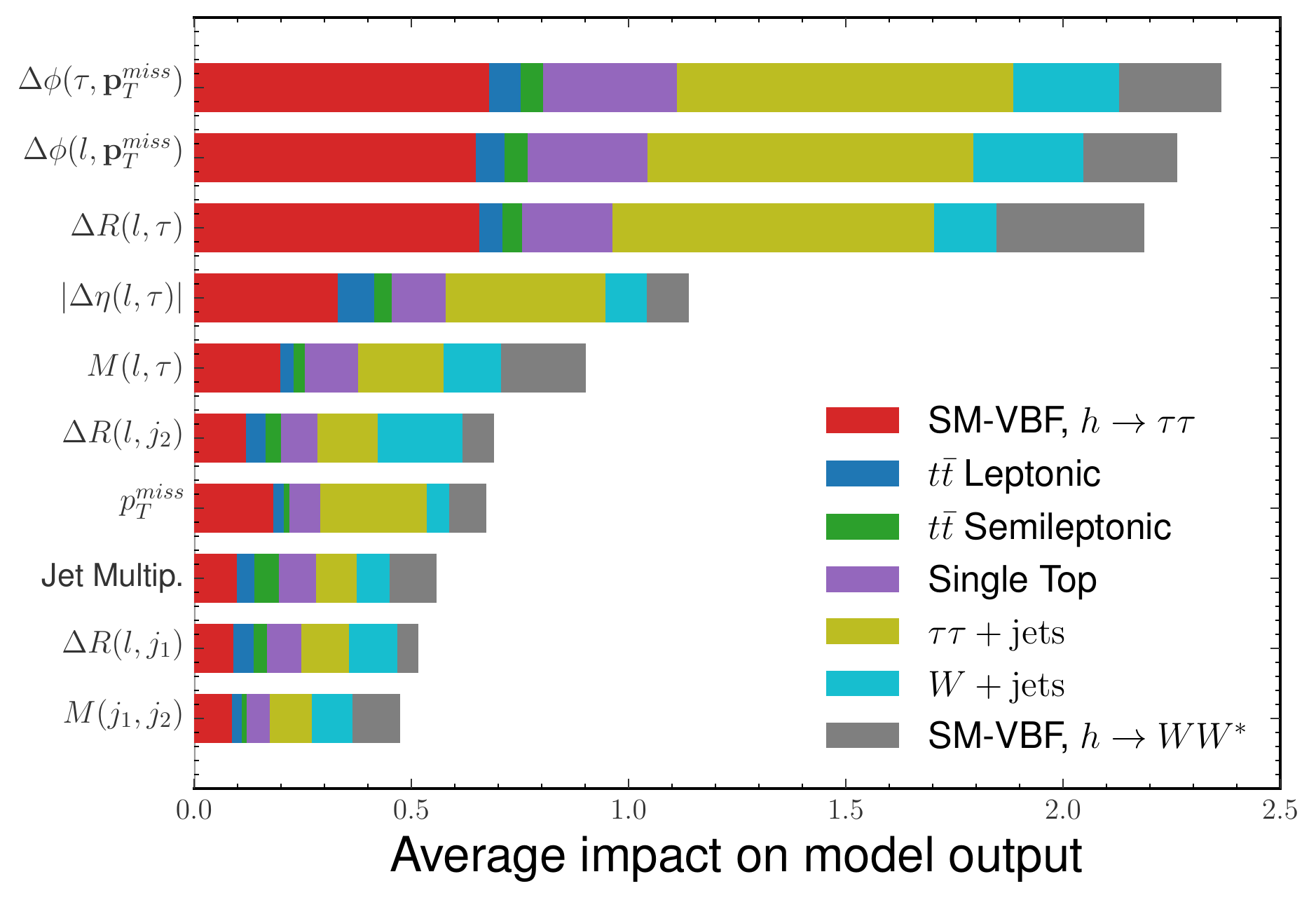} 
	\end{center}
	\caption{Same as Fig.~\ref{fig:diphoton_out} for semileptonic channel.}\label{fig:tLtH_out}
\end{figure}

\begin{figure}
	\begin{center}
		\includegraphics[scale=.35]{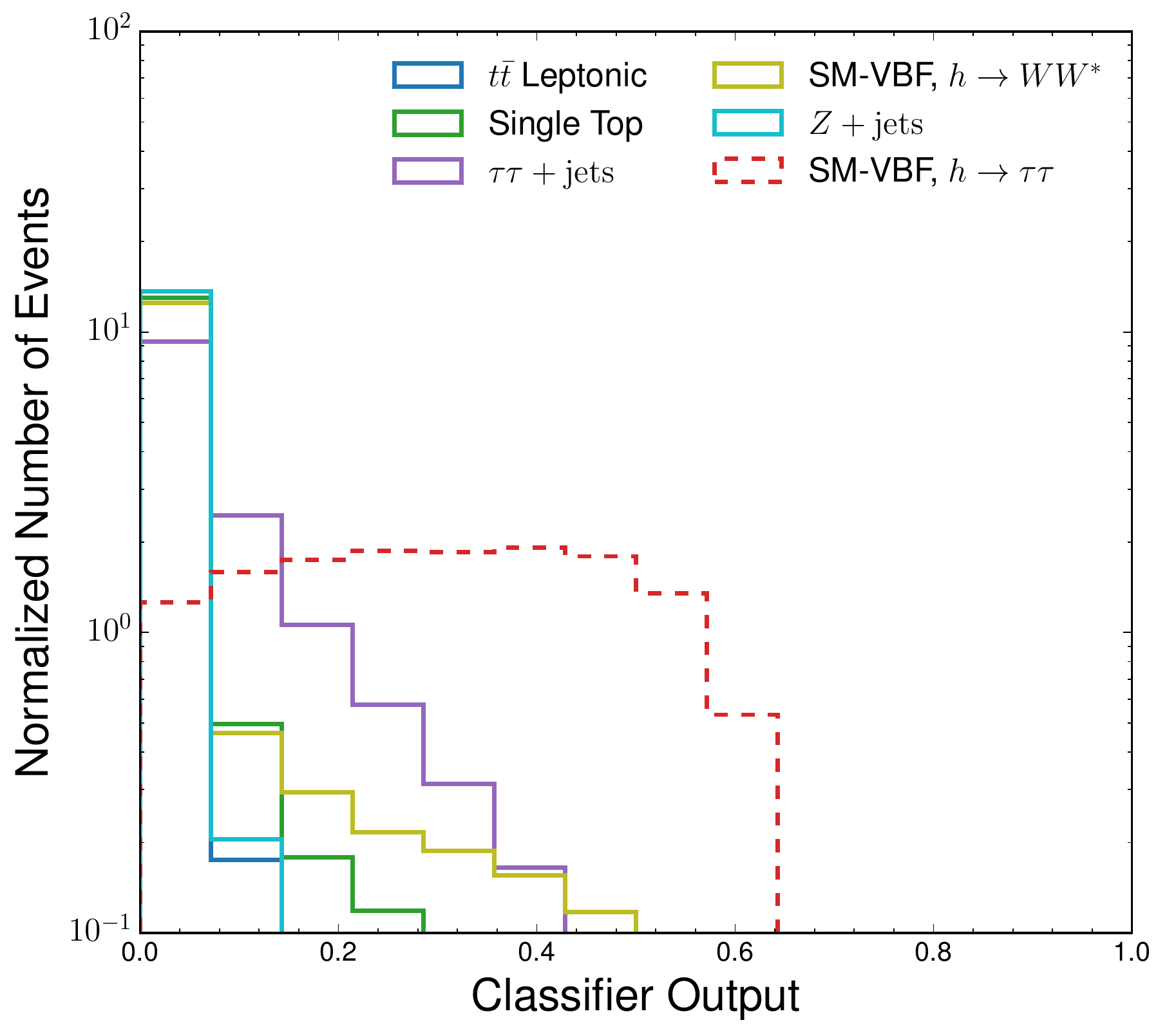}  
		\includegraphics[scale=.4]{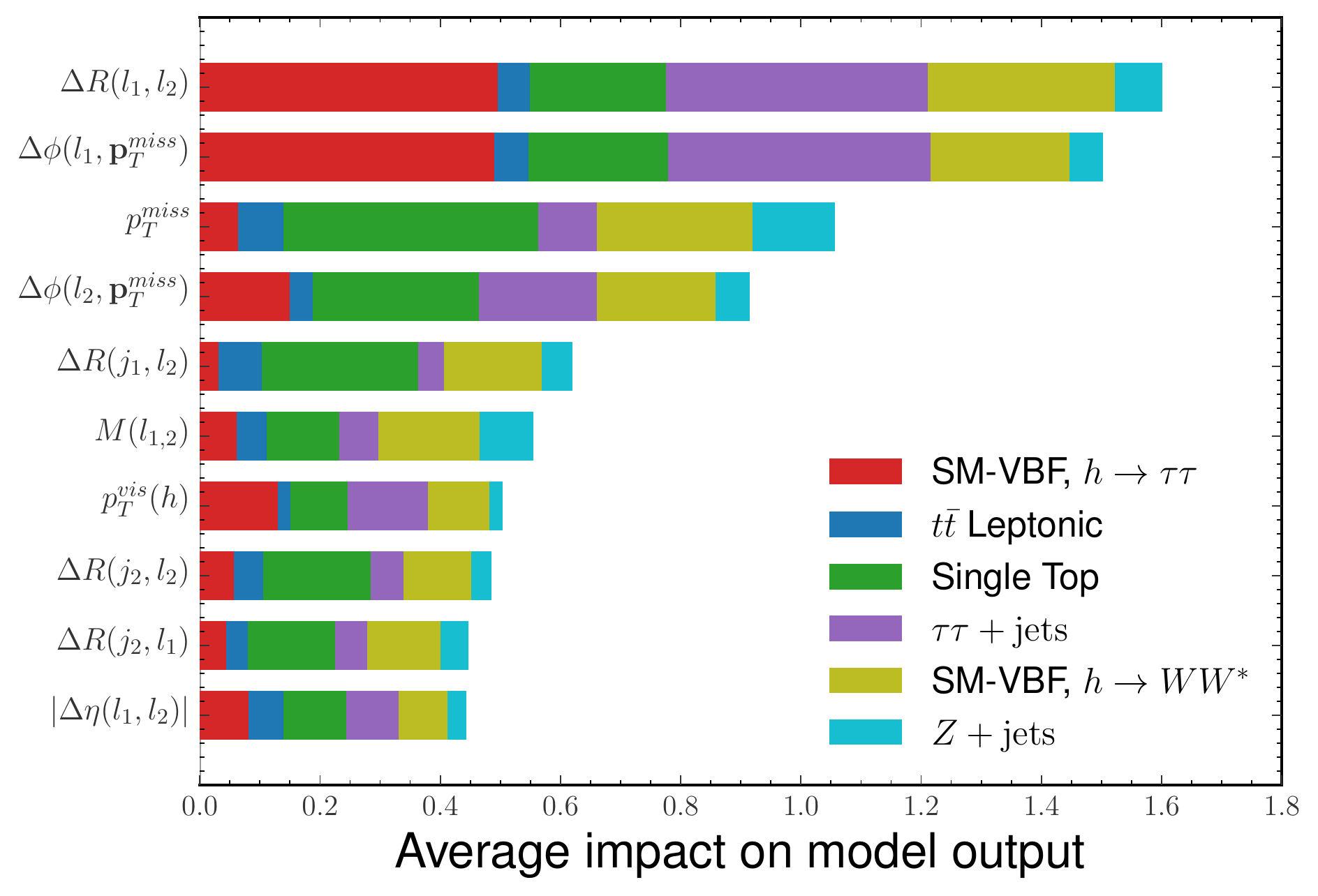} 
	\end{center}
	\caption{Same as Fig.~\ref{fig:diphoton_out} for leptonic channel.}\label{fig:tLtL_out}
\end{figure}

It is important to understand how the neural network learns and interprets the data, where understanding such features can help in optimising the cut-based analyses as well. For this reason, we adapt the {\sc SHapley Additive exPlanations (SHAP)}~\cite{2017arXiv170507874L} method. The {\sc SHAP} value shows the average of the marginal contributions of the input parameters to the neural network. In order to measure this value, we use the same training and test samples where the {\sc SHAP} explainer trained with 2000 events from the training sample and the {\sc SHAP} values are extracted using 1000 random events from the test sample. For diphoton channel, the most important ten observables with {\sc SHAP} values are presented in the right panel of Fig.~\ref{fig:diphoton_out} where the signal values are represented with red and the background values are represented with blue bars where the average {\sc SHAP} value has been divided by the contributions coming from the signal and the background. Although the {\sc SHAP} values are relatively low, one can immediately see the importance of the angular observables and transverse momenta of the second leading photon in the NN. The left panel of Fig.~\ref{fig:diphoton_out} shows the classifier output for the NN architecture where the red line shows the signal and the blue bars shows the background sample.
In figures~\ref{fig:tHtH_out}, \ref{fig:tLtH_out} and \ref{fig:tLtL_out}, we show the classifier output (on the left) and the {\sc SHAP} values for the ten parameters that have the biggest impact on the classification for hadronic, semi-leptonic and leptonic final states respectively. As seen in the {\sc SHAP} values, the ditau signal regions mostly rely on angular observables between final state particles. One can see in the classifier outputs, $ \tau\tau + $jets is the most dominant background. The loss of sensitivity can also be observed in the classifier outputs where the leptonic signal region can not reach beyond 70\%. This outcome renders semi-leptonic and leptonic signal regions as not optimal for sensitivity studies of the particular operator in hand.

In all, we observe much superior results in the diphoton signal region in terms of both statistical significance and accuracy of the NN. As presented in Appendix~\ref{app:RCBA}, compared to cut-based analysis our $S/B$ ratio is significantly lower in the NN analysis which is by design. We observed that by increasing yielding number of signal events one can populate the high energetic regions that are crucial for the sensitivity of EFT operators. We observe up to $38\%$ improvement depending on the choice of confidence level, luminosity and systematic uncertainty in the operator sensitivities due to large number of signal events left after the classification. This is because large statistical significance in the rectangular cut-based approach has been achieved with less yielding events which degrades the impact of the events where EFT effects are most prominent. In the ditau signal regions, due to the vast amount of background sources, the classification accuracy is lower than the diphoton channel. Expectedly, the $\tau\tau +$ jets background is the most dominant background source for all ditau subregions. All these results also compared with a boosted decision tree (BDT) algorithm. Although the BDT results were slightly less significant compared to the NN, we observed that both methods were giving priorities to similar observables (as represented by average SHAP values in NN case) to increase signal significance.

\section{Contribution from other anomalous couplings and other processes}
\label{subdominant}

In Sec.~\ref{sec:theory}, we argued why the contact terms, $g^h_{Zf}$, dominate at high energies if we assume a similar cut-off for all the couplings so that all the different anomalous couplings have a similar size $\sim v^2/\Lambda^2$. In this appendix we show that the dominance of the linear combination $g^{h(VBF)}_{Zf}$ holds even if we let all the anomalous couplings saturate their bounds; this is not obvious as these bounds are not all of  a similar size. Consider first the couplings, $\kappa_{WW}$ and $\kappa_{ZZ}$, which also generate a contribution that grows with ${t}$, albeit not as rapidly as the $g^h_{Zf}$ contributions.  Amongst these two couplings  $\kappa_{WW}$ has a much larger contribution to the process because of the  greater $W$-luminosity in the VBF process. This coupling  can be constrained using the following correlation that holds in the D6 SMEFT, 
\bea
\kappa_{WW}=\kappa_{\gamma\gamma}+\cot \theta_W \kappa_{Z\gamma}+\delta \kappa_{\gamma}
\eea
where the last two couplings are defined by the lagrangian terms, 
\bea
\Delta{\cal L}_6\supset \kappa_{\gamma\gamma}\,\frac{h}{2v} A^{\mu\nu}A_{\mu\nu}+\kappa_{Z\gamma}\,\frac{h}{v} A^{\mu\nu}{Z}_{\mu\nu}\,
\eea
The couplings $\kappa_{\gamma\gamma}, \kappa_{Z\gamma}$ can be already constrained at  the per-mille level or smaller.  On the other hand, even for the less constrained $\delta \kappa_{\gamma}$,   diboson processes could be used to obtain the strong constraint, 
\bea
|\delta \kappa_{\gamma}| \lesssim 0.005
\eea
at the HL-LHC~\cite{Grojean:2018dqj}. Taking a value of  $\delta \kappa_{\gamma}$  that saturates this  we find that, for the most sensitive bins around $p^h_T \sim 300$ GeV,  its interference contribution is 8 times smaller than that of $g^{h(VBF)}_{Zf}$ at its maximal value for the HL-LHC in \eq{fbound}. Thus the $g^{h(VBF)}_{Zf}$ contribution clearly dominates over the others.
 
 We now discuss the effect of the couplings,  $\kappa_{\gamma \gamma},\delta g^V_{f}, \delta \hat{g}^h_{\tau\tau}$ and $\delta \hat{g}^h_{VV}$,  that rescale the amplitude and thus only modify the total rate. Of these, $\kappa_{\gamma \gamma}$ and $\delta g^V_{f}$ are highly constrained, at per-mille level or smaller, respectively, from  $W, Z$ decays at LEP and the Higgs diphoton decay mode at the LHC~\cite{Pomarol:2013zra}; these couplings can therefore be completely neglected given that their contributions do not grow with energy unlike the contribution of  $g^{h(VBF)}_{Zf}$. The other two couplings,  $\delta \hat{g}^h_{\tau\tau}$ and $\delta \hat{g}^h_{VV}$ rescale all differential distributions by a constant factor $r = 1 + 2(\delta \hat{g}^h_{\tau\tau}+\delta \hat{g}^h_{VV})$. While $\delta \hat{g}^h_{VV}$ can be constrained much more stringently in other processes such as, $gg \to h \to 4 \ell$ and $pp \to Wh/Zh$, the VBF process considered in this paper is the most sensitive way to directly probe  $\delta \hat{g}^h_{\tau\tau}$ .The sensitivity to $\delta \hat{g}^h_{\tau\tau}$  however comes from lower bins with a much higher number of events,  unlike the bound on  $g^{h(VBF)}_{Zf}$ which arises from the bins around $p^h_T = 300$ GeV. Indeed less than  2-4 $\%$ of the events, depending on  the $\tau\tau$ decay mode in question,  lie in the region  $p^h_T > 300$ GeV so that the effect on the  bound on $\delta \hat{g}^h_{\tau\tau}$ in our set-up is completely negligible if these events are not considered. Thus $r$ and $g^{h(VBF)}_{Zf}$ can be independently measured by considering separately  the bins with $p^h_T$ less than or more than 300 GeV. If this procedure reveals a non-vanishing, $\delta \hat{g}^h_{\tau\tau}$, its effect can be subtracted from the higher bins in a straightforward way. \footnote{One should ideally perform a fit to simultaneously ascertain  the rescaled SM term proportional to $r$ as well as the the quadratically growing term proportional to $g^{h(VBF)}_{Zf}$, but for our present purposes the above procedure is sufficient.}
\begin{figure}[!h]
	\begin{center}
		\includegraphics[scale=.4]{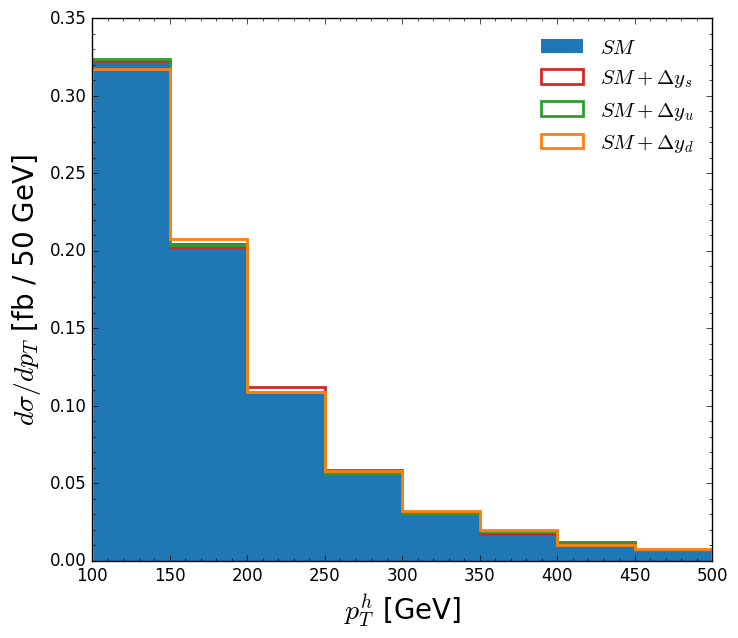}  
	\end{center}
	\caption{Figure shows the $p_T$ spectrum of the Higgs boson reconstructed from the diphoton channel. The blue, red, green and orange histograms show the contributions from the SM, SM + $\Delta y_s$, SM + $\Delta y_u$ and SM + $\Delta y_d$, respectively. $\Delta y_i$ ($i = s, u, d$) are the interferences with the SM for $\delta y_s \sim 17, \delta y_u \sim 700$ and $\delta y_d \sim 360$.}\label{fig:diphoton-yukawa-variation}
\end{figure}
Finally, we discuss how the variation of the Higgs Yukawas to the light quarks affect our results. We consider constraints from the Higgs signal strengths while allowing for enhanced effects from $q\bar{q} \to h$ production~\cite{Falkowski:2020znk, Delaunay:2016brc}. Following the former reference, we consider the modifications to the $u$, $d$ and $s$ Yukawa couplings to be $\delta y_d \lesssim 360, \delta y_u \lesssim 700$ and $\delta y_s \lesssim 17$. These bounds also include a future one on the signal strength to be $\mu = 1.00 \pm 0.03$ at the HL-LHC. Upon using these deformations, and our generation level cuts (as discussed above), we find the cross-sections to be modified as follows. For SM, $\delta y_d, \delta y_u$ and $\delta y_s$, the central values of the cross-sections are respectively 4.550 fb, 4.560 fb, 4.564 fb and 4.552 fb. We show the $p_T$ spectrum of the reconstructed Higgs boson in the diphoton channel in Fig.~\ref{fig:diphoton-yukawa-variation}. We find the modifications owing to the change in these Yukawas to be negligible.

\begin{acknowledgments}
	We thank Shilpi Jain and Sanmay Ganguly for several helpful discussions concerning the analyses. JYA acknowledges the funding received from the European Union's Horizon 2020 research and innovation programme as part of the Marie Skłodowska-Curie Innovative Training Network MCnetITN3 (grant agreement no. 722104). SB acknowledges the grant received from IPPP, Durham, where the majority of the work was done.
\end{acknowledgments}
	
	\bibliography{VBF_EFT}

\providecommand{\href}[2]{#2}\begingroup\raggedright\begin{thebibliography}{100}

\bibitem{Buchmuller:1985jz}
W.~Buchmuller and D.~Wyler, \emph{{Effective Lagrangian Analysis of New
  Interactions and Flavor Conservation}},
  \href{http://dx.doi.org/10.1016/0550-3213(86)90262-2}{\emph{Nucl. Phys.} {\bf
  B268} (1986) 621--653}.

\bibitem{Giudice:2007fh}
G.~F. Giudice, C.~Grojean, A.~Pomarol and R.~Rattazzi, \emph{{The
  Strongly-Interacting Light Higgs}},
  \href{http://dx.doi.org/10.1088/1126-6708/2007/06/045}{\emph{JHEP} {\bf 06}
  (2007) 045}, [\href{http://arxiv.org/abs/hep-ph/0703164}{{\tt
  hep-ph/0703164}}].

\bibitem{Grzadkowski:2010es}
B.~Grzadkowski, M.~Iskrzynski, M.~Misiak and J.~Rosiek, \emph{{Dimension-Six
  Terms in the Standard Model Lagrangian}},
  \href{http://dx.doi.org/10.1007/JHEP10(2010)085}{\emph{JHEP} {\bf 10} (2010)
  085}, [\href{http://arxiv.org/abs/1008.4884}{{\tt 1008.4884}}].

\bibitem{Gupta:2011be}
R.~S. Gupta, \emph{{Probing Quartic Neutral Gauge Boson Couplings using
  diffractive photon fusion at the LHC}},
  \href{http://dx.doi.org/10.1103/PhysRevD.85.014006}{\emph{Phys. Rev.} {\bf
  D85} (2012) 014006}, [\href{http://arxiv.org/abs/1111.3354}{{\tt
  1111.3354}}].

\bibitem{Gupta:2012mi}
R.~S. Gupta, H.~Rzehak and J.~D. Wells, \emph{{How well do we need to measure
  Higgs boson couplings?}},
  \href{http://dx.doi.org/10.1103/PhysRevD.86.095001}{\emph{Phys. Rev.} {\bf
  D86} (2012) 095001}, [\href{http://arxiv.org/abs/1206.3560}{{\tt
  1206.3560}}].

\bibitem{Banerjee:2012xc}
S.~Banerjee, S.~Mukhopadhyay and B.~Mukhopadhyaya, \emph{{New Higgs
  interactions and recent data from the LHC and the Tevatron}},
  \href{http://dx.doi.org/10.1007/JHEP10(2012)062}{\emph{JHEP} {\bf 10} (2012)
  062}, [\href{http://arxiv.org/abs/1207.3588}{{\tt 1207.3588}}].

\bibitem{Gupta:2012fy}
R.~S. Gupta, M.~Montull and F.~Riva, \emph{{SUSY Faces its Higgs Couplings}},
  \href{http://dx.doi.org/10.1007/JHEP04(2013)132}{\emph{JHEP} {\bf 04} (2013)
  132}, [\href{http://arxiv.org/abs/1212.5240}{{\tt 1212.5240}}].

\bibitem{Banerjee:2013apa}
S.~Banerjee, S.~Mukhopadhyay and B.~Mukhopadhyaya, \emph{{Higher dimensional
  operators and the LHC Higgs data: The role of modified kinematics}},
  \href{http://dx.doi.org/10.1103/PhysRevD.89.053010}{\emph{Phys. Rev.} {\bf
  D89} (2014) 053010}, [\href{http://arxiv.org/abs/1308.4860}{{\tt
  1308.4860}}].

\bibitem{Gupta:2013zza}
R.~S. Gupta, H.~Rzehak and J.~D. Wells, \emph{{How well do we need to measure
  the Higgs boson mass and self-coupling?}},
  \href{http://dx.doi.org/10.1103/PhysRevD.88.055024}{\emph{Phys. Rev.} {\bf
  D88} (2013) 055024}, [\href{http://arxiv.org/abs/1305.6397}{{\tt
  1305.6397}}].

\bibitem{Elias-Miro:2013eta}
J.~Elias-Mir{\'o}, C.~Grojean, R.~S. Gupta and D.~Marzocca, \emph{{Scaling and
  tuning of EW and Higgs observables}},
  \href{http://dx.doi.org/10.1007/JHEP05(2014)019}{\emph{JHEP} {\bf 05} (2014)
  019}, [\href{http://arxiv.org/abs/1312.2928}{{\tt 1312.2928}}].

\bibitem{Contino:2013kra}
R.~Contino, M.~Ghezzi, C.~Grojean, M.~Muhlleitner and M.~Spira,
  \emph{{Effective Lagrangian for a light Higgs-like scalar}},
  \href{http://dx.doi.org/10.1007/JHEP07(2013)035}{\emph{JHEP} {\bf 07} (2013)
  035}, [\href{http://arxiv.org/abs/1303.3876}{{\tt 1303.3876}}].

\bibitem{Falkowski:2014tna}
A.~Falkowski and F.~Riva, \emph{{Model-independent precision constraints on
  dimension-6 operators}},
  \href{http://dx.doi.org/10.1007/JHEP02(2015)039}{\emph{JHEP} {\bf 02} (2015)
  039}, [\href{http://arxiv.org/abs/1411.0669}{{\tt 1411.0669}}].

\bibitem{Englert:2014cva}
C.~Englert and M.~Spannowsky, \emph{{Effective Theories and Measurements at
  Colliders}},
  \href{http://dx.doi.org/10.1016/j.physletb.2014.11.035}{\emph{Phys. Lett.}
  {\bf B740} (2015) 8--15}, [\href{http://arxiv.org/abs/1408.5147}{{\tt
  1408.5147}}].

\bibitem{Gupta:2014rxa}
R.~S. Gupta, A.~Pomarol and F.~Riva, \emph{{BSM Primary Effects}},
  \href{http://dx.doi.org/10.1103/PhysRevD.91.035001}{\emph{Phys. Rev.} {\bf
  D91} (2015) 035001}, [\href{http://arxiv.org/abs/1405.0181}{{\tt
  1405.0181}}].

\bibitem{Amar:2014fpa}
G.~Amar, S.~Banerjee, S.~von Buddenbrock, A.~S. Cornell, T.~Mandal, B.~Mellado
  et~al., \emph{{Exploration of the tensor structure of the Higgs boson
  coupling to weak bosons in e$^{+}$ e$^{−}$ collisions}},
  \href{http://dx.doi.org/10.1007/JHEP02(2015)128}{\emph{JHEP} {\bf 02} (2015)
  128}, [\href{http://arxiv.org/abs/1405.3957}{{\tt 1405.3957}}].

\bibitem{Buschmann:2014sia}
M.~Buschmann, D.~Goncalves, S.~Kuttimalai, M.~Schonherr, F.~Krauss and
  T.~Plehn, \emph{{Mass Effects in the Higgs-Gluon Coupling: Boosted vs
  Off-Shell Production}},
  \href{http://dx.doi.org/10.1007/JHEP02(2015)038}{\emph{JHEP} {\bf 02} (2015)
  038}, [\href{http://arxiv.org/abs/1410.5806}{{\tt 1410.5806}}].

\bibitem{Craig:2014una}
N.~Craig, M.~Farina, M.~McCullough and M.~Perelstein, \emph{{Precision
  Higgsstrahlung as a Probe of New Physics}},
  \href{http://dx.doi.org/10.1007/JHEP03(2015)146}{\emph{JHEP} {\bf 03} (2015)
  146}, [\href{http://arxiv.org/abs/1411.0676}{{\tt 1411.0676}}].

\bibitem{Ellis:2014dva}
J.~Ellis, V.~Sanz and T.~You, \emph{{Complete Higgs Sector Constraints on
  Dimension-6 Operators}},
  \href{http://dx.doi.org/10.1007/JHEP07(2014)036}{\emph{JHEP} {\bf 07} (2014)
  036}, [\href{http://arxiv.org/abs/1404.3667}{{\tt 1404.3667}}].

\bibitem{Ellis:2014jta}
J.~Ellis, V.~Sanz and T.~You, \emph{{The Effective Standard Model after LHC Run
  I}}, \href{http://dx.doi.org/10.1007/JHEP03(2015)157}{\emph{JHEP} {\bf 03}
  (2015) 157}, [\href{http://arxiv.org/abs/1410.7703}{{\tt 1410.7703}}].

\bibitem{Banerjee:2015bla}
S.~Banerjee, T.~Mandal, B.~Mellado and B.~Mukhopadhyaya, \emph{{Cornering
  dimension-6 $HVV$ interactions at high luminosity LHC: the role of event
  ratios}}, \href{http://dx.doi.org/10.1007/JHEP09(2015)057}{\emph{JHEP} {\bf
  09} (2015) 057}, [\href{http://arxiv.org/abs/1505.00226}{{\tt 1505.00226}}].

\bibitem{Englert:2015hrx}
C.~Englert, R.~Kogler, H.~Schulz and M.~Spannowsky, \emph{{Higgs coupling
  measurements at the LHC}},
  \href{http://dx.doi.org/10.1140/epjc/s10052-016-4227-1}{\emph{Eur. Phys. J.}
  {\bf C76} (2016) 393}, [\href{http://arxiv.org/abs/1511.05170}{{\tt
  1511.05170}}].

\bibitem{Ghosh:2015gpa}
D.~Ghosh, R.~S. Gupta and G.~Perez, \emph{{Is the Higgs Mechanism of Fermion
  Mass Generation a Fact? A Yukawa-less First-Two-Generation Model}},
  \href{http://dx.doi.org/10.1016/j.physletb.2016.02.059}{\emph{Phys. Lett.}
  {\bf B755} (2016) 504--508}, [\href{http://arxiv.org/abs/1508.01501}{{\tt
  1508.01501}}].

\bibitem{Degrande:2016dqg}
C.~Degrande, B.~Fuks, K.~Mawatari, K.~Mimasu and V.~Sanz, \emph{{Electroweak
  Higgs boson production in the standard model effective field theory beyond
  leading order in QCD}},
  \href{http://dx.doi.org/10.1140/epjc/s10052-017-4793-x}{\emph{Eur. Phys. J.
  C} {\bf 77} (2017) 262}, [\href{http://arxiv.org/abs/1609.04833}{{\tt
  1609.04833}}].

\bibitem{Cohen:2016bsd}
J.~Cohen, S.~Bar-Shalom and G.~Eilam, \emph{{Contact Interactions in
  Higgs-Vector Boson Associated Production at the ILC}},
  \href{http://dx.doi.org/10.1103/PhysRevD.94.035030}{\emph{Phys. Rev.} {\bf
  D94} (2016) 035030}, [\href{http://arxiv.org/abs/1602.01698}{{\tt
  1602.01698}}].

\bibitem{Ge:2016zro}
S.-F. Ge, H.-J. He and R.-Q. Xiao, \emph{{Probing new physics scales from Higgs
  and electroweak observables at e$^{+}$ e$^{−}$ Higgs factory}},
  \href{http://dx.doi.org/10.1007/JHEP10(2016)007}{\emph{JHEP} {\bf 10} (2016)
  007}, [\href{http://arxiv.org/abs/1603.03385}{{\tt 1603.03385}}].

\bibitem{Contino:2016jqw}
R.~Contino, A.~Falkowski, F.~Goertz, C.~Grojean and F.~Riva, \emph{{On the
  Validity of the Effective Field Theory Approach to SM Precision Tests}},
  \href{http://dx.doi.org/10.1007/JHEP07(2016)144}{\emph{JHEP} {\bf 07} (2016)
  144}, [\href{http://arxiv.org/abs/1604.06444}{{\tt 1604.06444}}].

\bibitem{Biekotter:2016ecg}
A.~Biek{\"o}tter, J.~Brehmer and T.~Plehn, \emph{{Extending the limits of Higgs
  effective theory}},
  \href{http://dx.doi.org/10.1103/PhysRevD.94.055032}{\emph{Phys. Rev.} {\bf
  D94} (2016) 055032}, [\href{http://arxiv.org/abs/1602.05202}{{\tt
  1602.05202}}].

\bibitem{deBlas:2016ojx}
J.~de~Blas, M.~Ciuchini, E.~Franco, S.~Mishima, M.~Pierini, L.~Reina et~al.,
  \emph{{Electroweak precision observables and Higgs-boson signal strengths in
  the Standard Model and beyond: present and future}},
  \href{http://dx.doi.org/10.1007/JHEP12(2016)135}{\emph{JHEP} {\bf 12} (2016)
  135}, [\href{http://arxiv.org/abs/1608.01509}{{\tt 1608.01509}}].

\bibitem{Denizli:2017pyu}
H.~Denizli and A.~Senol, \emph{{Constraints on Higgs effective couplings in
  $H\nu \bar{\nu}$ production of CLIC at 380 GeV}},
  \href{http://dx.doi.org/10.1155/2018/1627051}{\emph{Adv. High Energy Phys.}
  {\bf 2018} (2018) 1627051}, [\href{http://arxiv.org/abs/1707.03890}{{\tt
  1707.03890}}].

\bibitem{Barklow:2017suo}
T.~Barklow, K.~Fujii, S.~Jung, R.~Karl, J.~List, T.~Ogawa et~al.,
  \emph{{Improved Formalism for Precision Higgs Coupling Fits}},
  \href{http://dx.doi.org/10.1103/PhysRevD.97.053003}{\emph{Phys. Rev.} {\bf
  D97} (2018) 053003}, [\href{http://arxiv.org/abs/1708.08912}{{\tt
  1708.08912}}].

\bibitem{Brivio:2017vri}
I.~Brivio and M.~Trott, \emph{{The Standard Model as an Effective Field
  Theory}}, \href{http://dx.doi.org/10.1016/j.physrep.2018.11.002}{\emph{Phys.
  Rept.} {\bf 793} (2019) 1--98}, [\href{http://arxiv.org/abs/1706.08945}{{\tt
  1706.08945}}].

\bibitem{Barklow:2017awn}
T.~Barklow, K.~Fujii, S.~Jung, M.~E. Peskin and J.~Tian,
  \emph{{Model-Independent Determination of the Triple Higgs Coupling at e+e-
  Colliders}}, \href{http://dx.doi.org/10.1103/PhysRevD.97.053004}{\emph{Phys.
  Rev.} {\bf D97} (2018) 053004}, [\href{http://arxiv.org/abs/1708.09079}{{\tt
  1708.09079}}].

\bibitem{Khanpour:2017cfq}
H.~Khanpour and M.~Mohammadi~Najafabadi, \emph{{Constraining Higgs boson
  effective couplings at electron-positron colliders}},
  \href{http://dx.doi.org/10.1103/PhysRevD.95.055026}{\emph{Phys. Rev.} {\bf
  D95} (2017) 055026}, [\href{http://arxiv.org/abs/1702.00951}{{\tt
  1702.00951}}].

\bibitem{Englert:2017aqb}
C.~Englert, R.~Kogler, H.~Schulz and M.~Spannowsky, \emph{{Higgs
  characterisation in the presence of theoretical uncertainties and invisible
  decays}}, \href{http://dx.doi.org/10.1140/epjc/s10052-017-5366-8}{\emph{Eur.
  Phys. J.} {\bf C77} (2017) 789}, [\href{http://arxiv.org/abs/1708.06355}{{\tt
  1708.06355}}].

\bibitem{panico}
G.~Panico, F.~Riva and A.~Wulzer, \emph{{Diboson Interference Resurrection}},
  \href{http://dx.doi.org/10.1016/j.physletb.2017.11.068}{\emph{Phys. Lett.}
  {\bf B776} (2018) 473--480}, [\href{http://arxiv.org/abs/1708.07823}{{\tt
  1708.07823}}].

\bibitem{Franceschini:2017xkh}
R.~Franceschini, G.~Panico, A.~Pomarol, F.~Riva and A.~Wulzer,
  \emph{{Electroweak Precision Tests in High-Energy Diboson Processes}},
  \href{http://dx.doi.org/10.1007/JHEP02(2018)111}{\emph{JHEP} {\bf 02} (2018)
  111}, [\href{http://arxiv.org/abs/1712.01310}{{\tt 1712.01310}}].

\bibitem{banerjee1}
S.~Banerjee, C.~Englert, R.~S. Gupta and M.~Spannowsky, \emph{{Probing
  Electroweak Precision Physics via boosted Higgs-strahlung at the LHC}},
  \href{http://dx.doi.org/10.1103/PhysRevD.98.095012}{\emph{Phys. Rev. D} {\bf
  98} (2018) 095012}, [\href{http://arxiv.org/abs/1807.01796}{{\tt
  1807.01796}}].

\bibitem{Grojean:2018dqj}
C.~Grojean, M.~Montull and M.~Riembau, \emph{Diboson at the lhc vs lep},
  \href{http://dx.doi.org/10.1007/JHEP03(2019)020}{\emph{JHEP} {\bf 03} (2019)
  020}, [\href{http://arxiv.org/abs/1810.05149}{{\tt 1810.05149}}].

\bibitem{Biekotter:2018rhp}
A.~Biekoetter, T.~Corbett and T.~Plehn, \emph{{The Gauge-Higgs Legacy of the
  LHC Run II}},
  \href{http://dx.doi.org/10.21468/SciPostPhys.6.6.064}{\emph{SciPost Phys.}
  {\bf 6} (2019) 064}, [\href{http://arxiv.org/abs/1812.07587}{{\tt
  1812.07587}}].

\bibitem{Goncalves:2018ptp}
D.~Goncalves and J.~Nakamura, \emph{{Boosting the $H\to$ invisibles searches
  with $Z$ boson polarization}},
  \href{http://dx.doi.org/10.1103/PhysRevD.99.055021}{\emph{Phys. Rev.} {\bf
  D99} (2019) 055021}, [\href{http://arxiv.org/abs/1809.07327}{{\tt
  1809.07327}}].

\bibitem{Gomez-Ambrosio:2018pnl}
R.~Gomez-Ambrosio, \emph{{Studies of Dimension-Six EFT effects in Vector Boson
  Scattering}},
  \href{http://dx.doi.org/10.1140/epjc/s10052-019-6893-2}{\emph{Eur. Phys. J.
  C} {\bf 79} (2019) 389}, [\href{http://arxiv.org/abs/1809.04189}{{\tt
  1809.04189}}].

\bibitem{Freitas:2019hbk}
F.~F. Freitas, C.~K. Khosa and V.~Sanz, \emph{{Exploring SMEFT in VH with
  Machine Learning}},  \href{http://arxiv.org/abs/1902.05803}{{\tt
  1902.05803}}.

\bibitem{Banerjee:2019pks}
S.~Banerjee, R.~S. Gupta, J.~Y. Reiness and M.~Spannowsky, \emph{{Resolving the
  tensor structure of the Higgs coupling to $Z$-bosons via Higgs-strahlung}},
  \href{http://dx.doi.org/10.1103/PhysRevD.100.115004}{\emph{Phys. Rev.} {\bf
  D100} (2019) 115004}, [\href{http://arxiv.org/abs/1905.02728}{{\tt
  1905.02728}}].

\bibitem{Banerjee:2019twi}
S.~Banerjee, R.~S. Gupta, J.~Y. Reiness, S.~Seth and M.~Spannowsky,
  \emph{{Towards the ultimate differential SMEFT analysis}},
  \href{http://dx.doi.org/10.1007/JHEP09(2020)170}{\emph{JHEP} {\bf 09} (2020)
  170}, [\href{http://arxiv.org/abs/1912.07628}{{\tt 1912.07628}}].

\bibitem{Biekotter:2020flu}
A.~Biek\"otter, R.~Gomez-Ambrosio, P.~Gregg, F.~Krauss and M.~Sch\"onherr,
  \emph{{Constraining SMEFT operators with associated $h\gamma$ production in
  Weak Boson Fusion}},  \href{http://arxiv.org/abs/2003.06379}{{\tt
  2003.06379}}.

\bibitem{Liu:2018pkg}
D.~Liu and L.-T. Wang, \emph{{Prospects for precision measurement of diboson
  processes in the semileptonic decay channel in future LHC runs}},
  \href{http://dx.doi.org/10.1103/PhysRevD.99.055001}{\emph{Phys. Rev. D} {\bf
  99} (2019) 055001}, [\href{http://arxiv.org/abs/1804.08688}{{\tt
  1804.08688}}].

\bibitem{Binosi:2008ig}
D.~Binosi, J.~Collins, C.~Kaufhold and L.~Theussl, \emph{{JaxoDraw: A Graphical
  user interface for drawing Feynman diagrams. Version 2.0 release notes}},
  \href{http://dx.doi.org/10.1016/j.cpc.2009.02.020}{\emph{Comput. Phys.
  Commun.} {\bf 180} (2009) 1709--1715},
  [\href{http://arxiv.org/abs/0811.4113}{{\tt 0811.4113}}].

\bibitem{Pomarol:2014dya}
A.~Pomarol, \emph{{Higgs Physics}},  in \emph{{2014 European School of
  High-Energy Physics}}, pp.~59--77, 2016.
\newblock \href{http://arxiv.org/abs/1412.4410}{{\tt 1412.4410}}.

\bibitem{DAmbrosio:2002vsn}
G.~D'Ambrosio, G.~Giudice, G.~Isidori and A.~Strumia, \emph{{Minimal flavor
  violation: An Effective field theory approach}},
  \href{http://dx.doi.org/10.1016/S0550-3213(02)00836-2}{\emph{Nucl. Phys. B}
  {\bf 645} (2002) 155--187}, [\href{http://arxiv.org/abs/hep-ph/0207036}{{\tt
  hep-ph/0207036}}].

\bibitem{Hagiwara:1986vm}
K.~Hagiwara, R.~D. Peccei, D.~Zeppenfeld and K.~Hikasa, \emph{{Probing the Weak
  Boson Sector in e+ e- ---> W+ W-}},
  \href{http://dx.doi.org/10.1016/0550-3213(87)90685-7}{\emph{Nucl. Phys.} {\bf
  B282} (1987) 253--307}.

\bibitem{Peskin:1991sw}
M.~E. Peskin and T.~Takeuchi, \emph{{Estimation of oblique electroweak
  corrections}}, \href{http://dx.doi.org/10.1103/PhysRevD.46.381}{\emph{Phys.
  Rev.} {\bf D46} (1992) 381--409}.

\bibitem{Barbieri:2004qk}
R.~Barbieri, A.~Pomarol, R.~Rattazzi and A.~Strumia, \emph{{Electroweak
  symmetry breaking after LEP-1 and LEP-2}},
  \href{http://dx.doi.org/10.1016/j.nuclphysb.2004.10.014}{\emph{Nucl. Phys.}
  {\bf B703} (2004) 127--146}, [\href{http://arxiv.org/abs/hep-ph/0405040}{{\tt
  hep-ph/0405040}}].

\bibitem{Wells:2015uba}
J.~D. Wells and Z.~Zhang, \emph{{Effective theories of universal theories}},
  \href{http://dx.doi.org/10.1007/JHEP01(2016)123}{\emph{JHEP} {\bf 01} (2016)
  123}, [\href{http://arxiv.org/abs/1510.08462}{{\tt 1510.08462}}].

\bibitem{Aaboud:2018pen}
{\scshape ATLAS} collaboration, M.~Aaboud et~al., \emph{{Cross-section
  measurements of the Higgs boson decaying into a pair of $\tau$-leptons in
  proton-proton collisions at $\sqrt{s}=13$ TeV with the ATLAS detector}},
  \href{http://dx.doi.org/10.1103/PhysRevD.99.072001}{\emph{Phys. Rev.} {\bf
  D99} (2019) 072001}, [\href{http://arxiv.org/abs/1811.08856}{{\tt
  1811.08856}}].

\bibitem{Cacciari:2008gp}
M.~Cacciari, G.~P. Salam and G.~Soyez, \emph{{The Anti-k(t) jet clustering
  algorithm}},
  \href{http://dx.doi.org/10.1088/1126-6708/2008/04/063}{\emph{JHEP} {\bf 04}
  (2008) 063}, [\href{http://arxiv.org/abs/0802.1189}{{\tt 0802.1189}}].

\bibitem{Alloul:2013bka}
A.~Alloul, N.~D. Christensen, C.~Degrande, C.~Duhr and B.~Fuks,
  \emph{{FeynRules 2.0 - A complete toolbox for tree-level phenomenology}},
  \href{http://dx.doi.org/10.1016/j.cpc.2014.04.012}{\emph{Comput. Phys.
  Commun.} {\bf 185} (2014) 2250--2300},
  [\href{http://arxiv.org/abs/1310.1921}{{\tt 1310.1921}}].

\bibitem{Degrande:2011ua}
C.~Degrande, C.~Duhr, B.~Fuks, D.~Grellscheid, O.~Mattelaer and T.~Reiter,
  \emph{Ufo - the universal feynrules output},
  \href{http://dx.doi.org/10.1016/j.cpc.2012.01.022}{\emph{Comput. Phys.
  Commun.} {\bf 183} (2012) 1201--1214},
  [\href{http://arxiv.org/abs/1108.2040}{{\tt 1108.2040}}].

\bibitem{Alwall:2014hca}
J.~Alwall, R.~Frederix, S.~Frixione, V.~Hirschi, F.~Maltoni, O.~Mattelaer
  et~al., \emph{{The automated computation of tree-level and next-to-leading
  order differential cross sections, and their matching to parton shower
  simulations}}, \href{http://dx.doi.org/10.1007/JHEP07(2014)079}{\emph{JHEP}
  {\bf 07} (2014) 079}, [\href{http://arxiv.org/abs/1405.0301}{{\tt
  1405.0301}}].

\bibitem{Sjostrand:2014zea}
T.~Sj{\"o}strand, S.~Ask, J.~R. Christiansen, R.~Corke, N.~Desai, P.~Ilten
  et~al., \emph{{An Introduction to PYTHIA 8.2}},
  \href{http://dx.doi.org/10.1016/j.cpc.2015.01.024}{\emph{Comput. Phys.
  Commun.} {\bf 191} (2015) 159--177},
  [\href{http://arxiv.org/abs/1410.3012}{{\tt 1410.3012}}].

\bibitem{Ball:2014uwa}
{\scshape NNPDF} collaboration, R.~D. Ball et~al., \emph{{Parton distributions
  for the LHC Run II}},
  \href{http://dx.doi.org/10.1007/JHEP04(2015)040}{\emph{JHEP} {\bf 04} (2015)
  040}, [\href{http://arxiv.org/abs/1410.8849}{{\tt 1410.8849}}].

\bibitem{Buckley:2014ana}
A.~Buckley, J.~Ferrando, S.~Lloyd, K.~Nordstr{\"o}m, B.~Page, M.~R{\"u}fenacht
  et~al., \emph{{LHAPDF6: parton density access in the LHC precision era}},
  \href{http://dx.doi.org/10.1140/epjc/s10052-015-3318-8}{\emph{Eur. Phys. J.}
  {\bf C75} (2015) 132}, [\href{http://arxiv.org/abs/1412.7420}{{\tt
  1412.7420}}].

\bibitem{Greljo:2017spw}
A.~Greljo, G.~Isidori, J.~M. Lindert, D.~Marzocca and H.~Zhang,
  \emph{{Electroweak Higgs production with HiggsPO at NLO QCD}},
  \href{http://dx.doi.org/10.1140/epjc/s10052-017-5422-4}{\emph{Eur. Phys. J.
  C} {\bf 77} (2017) 838}, [\href{http://arxiv.org/abs/1710.04143}{{\tt
  1710.04143}}].

\bibitem{Kallweit:2015dum}
S.~Kallweit, J.~M. Lindert, P.~Maierhofer, S.~Pozzorini and M.~Sch\"onherr,
  \emph{{NLO QCD+EW predictions for V + jets including off-shell vector-boson
  decays and multijet merging}},
  \href{http://dx.doi.org/10.1007/JHEP04(2016)021}{\emph{JHEP} {\bf 04} (2016)
  021}, [\href{http://arxiv.org/abs/1511.08692}{{\tt 1511.08692}}].

\bibitem{twiki}
{\url{https://twiki.cern.ch/twiki/bin/view/LHCPhysics/TtbarNNLO}}.

\bibitem{Kant:2014oha}
P.~Kant, O.~Kind, T.~Kintscher, T.~Lohse, T.~Martini, S.~M\"olbitz et~al.,
  \emph{{HatHor for single top-quark production: Updated predictions and
  uncertainty estimates for single top-quark production in hadronic
  collisions}},
  \href{http://dx.doi.org/10.1016/j.cpc.2015.02.001}{\emph{Comput. Phys.
  Commun.} {\bf 191} (2015) 74--89},
  [\href{http://arxiv.org/abs/1406.4403}{{\tt 1406.4403}}].

\bibitem{Elagin:2010aw}
A.~Elagin, P.~Murat, A.~Pranko and A.~Safonov, \emph{{A New Mass Reconstruction
  Technique for Resonances Decaying to di-tau}},
  \href{http://dx.doi.org/10.1016/j.nima.2011.07.009}{\emph{Nucl. Instrum.
  Meth. A} {\bf 654} (2011) 481--489},
  [\href{http://arxiv.org/abs/1012.4686}{{\tt 1012.4686}}].

\bibitem{Aaboud:2018xdt}
{\scshape ATLAS} collaboration, M.~Aaboud et~al., \emph{{Measurements of Higgs
  boson properties in the diphoton decay channel with 36 fb$^{-1}$ of $pp$
  collision data at $\sqrt{s} = 13$ TeV with the ATLAS detector}},
  \href{http://dx.doi.org/10.1103/PhysRevD.98.052005}{\emph{Phys. Rev. D} {\bf
  98} (2018) 052005}, [\href{http://arxiv.org/abs/1802.04146}{{\tt
  1802.04146}}].

\bibitem{ATLAS-CONF-2014-018}
\emph{{Tagging and suppression of pileup jets with the ATLAS detector}},  Tech.
  Rep. ATLAS-CONF-2014-018, CERN, Geneva, May, 2014.

\bibitem{Aad:2013aa}
G.~Aad, T.~Abajyan, B.~Abbott, J.~Abdallah, S.~A. Khalek, A.~A. Abdelalim
  et~al., \emph{Measurement of isolated-photon pair production in pp collisions
  at $ \sqrt{s} = 7 $ tev with the atlas detector},
  \href{http://dx.doi.org/10.1007/JHEP01(2013)086}{\emph{Journal of High Energy
  Physics} {\bf 2013} (2013) 86}.

\bibitem{PhysRevD.85.012003}
{\scshape ATLAS Collaboration} collaboration, G.~Aad, B.~Abbott, J.~Abdallah,
  A.~A. Abdelalim, A.~Abdesselam, O.~Abdinov et~al., \emph{Measurement of the
  isolated diphoton cross section in $pp$ collisions at
  $\sqrt{s}\mathbf{=}7\text{ }\text{ }\mathrm{TeV}$ with the atlas detector},
  \href{http://dx.doi.org/10.1103/PhysRevD.85.012003}{\emph{Phys. Rev. D} {\bf
  85} (Jan, 2012) 012003}.

\bibitem{Conte:2018vmg}
E.~Conte and B.~Fuks, \emph{{Confronting new physics theories to LHC data with
  MADANALYSIS 5}},
  \href{http://dx.doi.org/10.1142/S0217751X18300272}{\emph{Int. J. Mod. Phys.}
  {\bf A33} (2018) 1830027}, [\href{http://arxiv.org/abs/1808.00480}{{\tt
  1808.00480}}].

\bibitem{Cacciari:2011ma}
M.~Cacciari, G.~P. Salam and G.~Soyez, \emph{{FastJet User Manual}},
  \href{http://dx.doi.org/10.1140/epjc/s10052-012-1896-2}{\emph{Eur. Phys. J.}
  {\bf C72} (2012) 1896}, [\href{http://arxiv.org/abs/1111.6097}{{\tt
  1111.6097}}].

\bibitem{LEP2}
{\scshape ALEPH Collaboration, DELPHI Collaboration, L3 Collaboration, OPAL
  Collaboration, LEP TGC Working Group} collaboration, \emph{{A Combination of
  Preliminary Results on Gauge Boson Couplings Measured by the LEP
  experiments}},  Tech. Rep. LEPEWWG-TGC-2003-01. DELPHI-2003-068-PHYS-936.
  L3-Note-2826. LEPEWWG-2006-01. OPAL-TN-739. ALEPH-2006-016-CONF-2003-012,
  CERN, Geneva, Jan, 2003.

\bibitem{Baak:2012kk}
M.~Baak, M.~Goebel, J.~Haller, A.~Hoecker, D.~Kennedy, R.~Kogler et~al.,
  \emph{{The Electroweak Fit of the Standard Model after the Discovery of a New
  Boson at the LHC}},
  \href{http://dx.doi.org/10.1140/epjc/s10052-012-2205-9}{\emph{Eur. Phys. J.}
  {\bf C72} (2012) 2205}, [\href{http://arxiv.org/abs/1209.2716}{{\tt
  1209.2716}}].

\bibitem{Farina:2016ws}
M.~Farina, G.~Panico, D.~Pappadopulo, J.~T. Ruderman, R.~Torre and A.~Wulzer,
  \emph{{Energy helps accuracy: electroweak precision tests at hadron
  colliders}},
  \href{http://dx.doi.org/10.1016/j.physletb.2017.06.043}{\emph{Phys. Lett. B}
  {\bf 772} (2017) 210--215}, [\href{http://arxiv.org/abs/1609.08157}{{\tt
  1609.08157}}].

\bibitem{Araz:2020lnp}
J.~Y. Araz, B.~Fuks and G.~Polykratis, \emph{{Simplified fast detector
  simulation in MadAnalysis 5}},  \href{http://arxiv.org/abs/2006.09387}{{\tt
  2006.09387}}.

\bibitem{Baldi:2014pta}
P.~Baldi, P.~Sadowski and D.~Whiteson, \emph{{Enhanced Higgs Boson to
  $\tau^+\tau^-$ Search with Deep Learning}},
  \href{http://dx.doi.org/10.1103/PhysRevLett.114.111801}{\emph{Phys. Rev.
  Lett.} {\bf 114} (2015) 111801}, [\href{http://arxiv.org/abs/1410.3469}{{\tt
  1410.3469}}].

\bibitem{deOliveira:2015xxd}
L.~de~Oliveira, M.~Kagan, L.~Mackey, B.~Nachman and A.~Schwartzman,
  \emph{{Jet-images --- deep learning edition}},
  \href{http://dx.doi.org/10.1007/JHEP07(2016)069}{\emph{JHEP} {\bf 07} (2016)
  069}, [\href{http://arxiv.org/abs/1511.05190}{{\tt 1511.05190}}].

\bibitem{Baldi:2016fzo}
P.~Baldi, K.~Cranmer, T.~Faucett, P.~Sadowski and D.~Whiteson,
  \emph{{Parameterized neural networks for high-energy physics}},
  \href{http://dx.doi.org/10.1140/epjc/s10052-016-4099-4}{\emph{Eur. Phys. J.
  C} {\bf 76} (2016) 235}, [\href{http://arxiv.org/abs/1601.07913}{{\tt
  1601.07913}}].

\bibitem{Caron:2016hib}
S.~Caron, J.~S. Kim, K.~Rolbiecki, R.~Ruiz~de Austri and B.~Stienen, \emph{{The
  BSM-AI project: SUSY-AI--generalizing LHC limits on supersymmetry with
  machine learning}},
  \href{http://dx.doi.org/10.1140/epjc/s10052-017-4814-9}{\emph{Eur. Phys. J.
  C} {\bf 77} (2017) 257}, [\href{http://arxiv.org/abs/1605.02797}{{\tt
  1605.02797}}].

\bibitem{Chang:2017kvc}
S.~Chang, T.~Cohen and B.~Ostdiek, \emph{{What is the Machine Learning?}},
  \href{http://dx.doi.org/10.1103/PhysRevD.97.056009}{\emph{Phys. Rev. D} {\bf
  97} (2018) 056009}, [\href{http://arxiv.org/abs/1709.10106}{{\tt
  1709.10106}}].

\bibitem{Lin:2018cin}
J.~Lin, M.~Freytsis, I.~Moult and B.~Nachman, \emph{{Boosting $H\to b\bar b$
  with Machine Learning}},
  \href{http://dx.doi.org/10.1007/JHEP10(2018)101}{\emph{JHEP} {\bf 10} (2018)
  101}, [\href{http://arxiv.org/abs/1807.10768}{{\tt 1807.10768}}].

\bibitem{Albertsson:2018maf}
K.~Albertsson et~al., \emph{{Machine Learning in High Energy Physics Community
  White Paper}},
  \href{http://dx.doi.org/10.1088/1742-6596/1085/2/022008}{\emph{J. Phys. Conf.
  Ser.} {\bf 1085} (2018) 022008}, [\href{http://arxiv.org/abs/1807.02876}{{\tt
  1807.02876}}].

\bibitem{Guest:2018yhq}
D.~Guest, K.~Cranmer and D.~Whiteson, \emph{{Deep Learning and its Application
  to LHC Physics}},
  \href{http://dx.doi.org/10.1146/annurev-nucl-101917-021019}{\emph{Ann. Rev.
  Nucl. Part. Sci.} {\bf 68} (2018) 161--181},
  [\href{http://arxiv.org/abs/1806.11484}{{\tt 1806.11484}}].

\bibitem{Abdughani:2019wuv}
M.~Abdughani, J.~Ren, L.~Wu, J.~M. Yang and J.~Zhao, \emph{{Supervised deep
  learning in high energy phenomenology: a mini review}},
  \href{http://dx.doi.org/10.1088/0253-6102/71/8/955}{\emph{Commun. Theor.
  Phys.} {\bf 71} (2019) 955}, [\href{http://arxiv.org/abs/1905.06047}{{\tt
  1905.06047}}].

\bibitem{Windischhofer:2019ltt}
P.~Windischhofer, M.~Zgubi\v~c and D.~Bortoletto, \emph{{Preserving physically
  important variables in optimal event selections: A case study in Higgs
  physics}},  \href{http://arxiv.org/abs/1907.02098}{{\tt 1907.02098}}.

\bibitem{Amacker:2020bmn}
J.~Amacker et~al., \emph{{Higgs self-coupling measurements using deep learning
  and jet substructure in the $b\bar{b}b\bar{b}$ final state}},
  \href{http://arxiv.org/abs/2004.04240}{{\tt 2004.04240}}.

\bibitem{chollet2015keras}
F.~Chollet et~al., ``Keras.'' \url{https://keras.io}, 2015.

\bibitem{HORNIK1991251}
K.~Hornik, \emph{Approximation capabilities of multilayer feedforward
  networks},
  \href{http://dx.doi.org/https://doi.org/10.1016/0893-6080(91)90009-T}{\emph{Neural
  Networks} {\bf 4} (1991) 251 -- 257}.

\bibitem{1165576}
R.~{Lippmann}, \emph{An introduction to computing with neural nets},
  {\emph{IEEE ASSP Magazine} {\bf 4} (1987) 4--22}.

\bibitem{HORNIK1989359}
K.~Hornik, M.~Stinchcombe and H.~White, \emph{Multilayer feedforward networks
  are universal approximators},
  \href{http://dx.doi.org/https://doi.org/10.1016/0893-6080(89)90020-8}{\emph{Neural
  Networks} {\bf 2} (1989) 359 -- 366}.

\bibitem{Kingma2014AdamAM}
D.~P. Kingma and J.~Ba, \emph{Adam: A method for stochastic optimization},
  {\emph{CoRR} {\bf abs/1412.6980} (2014) }.

\bibitem{Aad:2019yxi}
{\scshape ATLAS} collaboration, G.~Aad et~al., \emph{{Search for non-resonant
  Higgs boson pair production in the $bb\ell\nu\ell\nu$ final state with the
  ATLAS detector in $pp$ collisions at $\sqrt{s} = 13$ TeV}},
  \href{http://dx.doi.org/10.1016/j.physletb.2019.135145}{\emph{Phys. Lett. B}
  {\bf 801} (2020) 135145}, [\href{http://arxiv.org/abs/1908.06765}{{\tt
  1908.06765}}].

\bibitem{L2Regularization}
C.~Cortes, M.~Mohri and A.~Rostamizadeh, \emph{$l_2$ regularization for
  learning kernels}, {\emph{CoRR} {\bf abs/1205.2653} (2012) },
  [\href{http://arxiv.org/abs/1205.2653}{{\tt 1205.2653}}].

\bibitem{Baringer:2011nh}
P.~Baringer, K.~Kong, M.~McCaskey and D.~Noonan, \emph{{Revisiting
  Combinatorial Ambiguities at Hadron Colliders with $M_{T2}$}},
  \href{http://dx.doi.org/10.1007/JHEP10(2011)101}{\emph{JHEP} {\bf 10} (2011)
  101}, [\href{http://arxiv.org/abs/1109.1563}{{\tt 1109.1563}}].

\bibitem{Barr:2013tda}
A.~J. Barr, M.~J. Dolan, C.~Englert and M.~Spannowsky, \emph{{Di-Higgs final
  states augMT2ed -- selecting $hh$ events at the high luminosity LHC}},
  \href{http://dx.doi.org/10.1016/j.physletb.2013.12.011}{\emph{Phys. Lett. B}
  {\bf 728} (2014) 308--313}, [\href{http://arxiv.org/abs/1309.6318}{{\tt
  1309.6318}}].

\bibitem{precision_recall}
J.~Davis and M.~Goadrich, \emph{The relationship between precision-recall and
  roc curves},  in \emph{Proceedings of the 23rd International Conference on
  Machine Learning}, ICML '06, (New York, NY, USA), pp.~233--240, Association
  for Computing Machinery, 2006.
\newblock \href{http://dx.doi.org/10.1145/1143844.1143874}{DOI}.

\bibitem{2017arXiv170507874L}
S.~{Lundberg} and S.-I. {Lee}, \emph{{A Unified Approach to Interpreting Model
  Predictions}}, {\emph{arXiv e-prints} (May, 2017) arXiv:1705.07874},
  [\href{http://arxiv.org/abs/1705.07874}{{\tt 1705.07874}}].

\bibitem{Pomarol:2013zra}
A.~Pomarol and F.~Riva, \emph{{Towards the Ultimate SM Fit to Close in on Higgs
  Physics}}, \href{http://dx.doi.org/10.1007/JHEP01(2014)151}{\emph{JHEP} {\bf
  01} (2014) 151}, [\href{http://arxiv.org/abs/1308.2803}{{\tt 1308.2803}}].

\bibitem{Falkowski:2020znk}
A.~Falkowski, S.~Ganguly, P.~Gras, J.~M. No, K.~Tobioka, N.~Vignaroli et~al.,
  \emph{{Light quark Yukawas in triboson final states}},
  \href{http://arxiv.org/abs/2011.09551}{{\tt 2011.09551}}.

\bibitem{Delaunay:2016brc}
C.~Delaunay, R.~Ozeri, G.~Perez and Y.~Soreq, \emph{{Probing Atomic Higgs-like
  Forces at the Precision Frontier}},
  \href{http://dx.doi.org/10.1103/PhysRevD.96.093001}{\emph{Phys. Rev. D} {\bf
  96} (2017) 093001}, [\href{http://arxiv.org/abs/1601.05087}{{\tt
  1601.05087}}].

\end{thebibliography}\endgroup
	
\end{document}